\documentclass[twocolumn,british,english,nofootinbib]{revtex4-2}
\usepackage[T1]{fontenc}
\usepackage[latin9]{inputenc}
\usepackage{color}
\definecolor{document_fontcolor}{rgb}{0, 0, 0}
\color{document_fontcolor}
\usepackage{babel}
\usepackage{float}
\usepackage{amsmath}
\usepackage{amssymb}
\usepackage{stmaryrd}
\usepackage{graphicx}
\usepackage[pdfusetitle,
 bookmarks=true,bookmarksnumbered=false,bookmarksopen=false,
 breaklinks=false,pdfborder={0 0 1},backref=false,colorlinks=true]
 {hyperref}

\makeatletter
\usepackage{dcolumn}
\usepackage{bm}
\usepackage{color}
\usepackage{soul}
\usepackage{babel}
\usepackage{amsfonts}
\usepackage{slashed}
\usepackage{enumerate}
\usepackage{orcidlink}

\makeatother

\begin{document}
\global\long\def\sgn{\mathrm{sgn}}%
\global\long\def\ket#1{\left|#1\right\rangle }%
\global\long\def\bra#1{\left\langle #1\right|}%
\global\long\def\sp#1#2{\langle#1|#2\rangle}%
\global\long\def\abs#1{\left|#1\right|}%
\global\long\def\avg#1{\langle#1\rangle}%

\title{On \foreignlanguage{british}{modeling} quantum point contacts in quantum
Hall systems}
\author{Prasoon Kumar\,\orcidlink{0009-0008-1564-5932}}
\affiliation{Univ. Grenoble Alpes, CEA, Grenoble INP, IRIG, PHELIQS, 38000 Grenoble,
France}
\author{Kyrylo Snizhko\,\orcidlink{0000-0002-7236-6779}}
\affiliation{Univ. Grenoble Alpes, CEA, Grenoble INP, IRIG, PHELIQS, 38000 Grenoble,
France}
\date{\today}
\begin{abstract}
Quantum point contacts (QPC) are a key instrument in investigating
the physics of edge excitations in the quantum Hall effect. However,
at not-so-high bias voltage values, the predictions of the conventional
point QPC model often deviate from the experimental data both in the
integer and (more prominently) in the fractional quantum Hall regime.
One of the possible explanations for such behaviors is the dependence
of the tunneling between the edges on energy, an effect not present
in the conventional model. Here we introduce two models that take
QPC spatial extension into account: wide-QPC model that accounts for
the distance along which the edges are in contact; long-QPC model
accounts for the fact that the tunneling amplitude originates from
a finite bulk gap and a finite distance between the two edges. We
investigate the predictions of these two models in the integer quantum
Hall regime for the energy dependence of the tunneling amplitude.
We find that these two models predict opposite dependences: the amplitude
decreasing or increasing away from the Fermi level. We thus elucidate
the effect of the QPC geometry on the energy dependence of the tunneling
amplitude and investigate its implications for transport observables.
\end{abstract}
\maketitle

\section{Introduction}

Quantum Hall (QH) effect is one of the most celebrated phenomena in
condensed matter physics \citep{Klitzing2017}. Beyond conductance
quantization, its fractional version exhibits extications with fractional
charge \citep{Laughlin1983,Goldman1995,FracChargeObs_Glattli,FracChargeObs_Heiblum,Dolev2008,Martin2004,Venkatachalam2011,Kapfer2019,Bisognin2019a}
and anyonic statistics \citep{Arovas1984,Nakamura2020,Bartolomei2020,Lee2022a,Glidic2023,Ruelle2023,Ruelle2025}.
Such unconventional quasiparticle excitations might be useful for
topological quantum computing \citep{Nayak2008}.

The origin of the conductance quantization is the formation of chiral
edge channels. These edge channels play a crucial role both in theoretical
understanding and experimental investigations of the quantum Hall
physics. Indeed, many theoretical predictions for the nature of quantum
Hall excitations stem from the chiral Luttinger liquid description
of those edge channels \citep{Wen2004}. A significant fraction of
quantum Hall experiments involves tunneling contacts at which quasiparticles
can jump between different edges --- quantum point contacts (QPCs)
\citep{FracChargeObs_Glattli,FracChargeObs_Heiblum,Dolev2008,Kapfer2019,Bisognin2019a,Nakamura2020,Bartolomei2020,Lee2022a,Glidic2023,Ruelle2023,Veillon2024,Ruelle2025,Guerrero-Suarez2025}.
Many advanced predictions have been verified through such experiments.

At the same time, experiments with a single QPC often show discrepancies
with theoretical predictions: they agree at small values of the bias
voltage, while at larger values ($eV$ larger than the system temperature,
$k_{\mathrm{B}}T$, yet significantly smaller than the bulk gap) they
show noticeable disagreement \citep{Schiller2024,Veillon2024,Guerrero-Suarez2025}.
With QPC being such an important instrument, it is crucial to understand
the origin of these discrepancies in order to understand the instrument's
limitations.

One of the possible explanations for the discrepancies is the energy
dependence of the tunneling amplitude. This dependence is generally
expected to be highly non-universal: it may depend on the QPC geometry,
impurities present in the sample etc. Here we attempt to investigate
the energy dependence of the tunneling amplitude from basic theoretical
considerations.

Figure~\ref{fig:1_QPC_experiment_scheme} shows the basic setup for
QPC experiments: The edges are in contact in a restricted region of
space. The tunneling is investigated by injecting current at source
contacts and measuring the output current and noise at the drain contacts.
In the conventional model, Fig.~\ref{fig:2_point_QPC_model}, the
tunneling is assumed to happen at a single point described by an energy-independent
tunneling amplitude. This point-QPC model is the standard model for
describing QPCs in the quantum Hall effect (see Refs.~\citep{Wen1991,Saleur2,LBFS_MachZehnder,Morel2021,Schiller2022a,Rosenow2025}
to mention but a few examples). The works that did consider realistic
geometries of the QPC \citep{Pryadko2000,Papa2004,Yang2013} focused
on the effect of Coulomb interactions and ignored non-point tunneling.

In this paper, we introduce two models that account for different
aspects of the realistic geometry of the QPC. The wide-QPC model shown
in Fig.~\ref{fig:wide-QPC_model} accounts for an extended width
of the QPC --- the distance along which the two edges are in contact
with each other. The long-QPC model shown in Fig.~\ref{fig:long-QPC_model}
accounts for the fact that the tunneling amplitude originates from
a finite bulk gap and for a finite distance between the two edges,
and models the effective tunneling barrier between the edges. The
long-QPC model essentially models a small piece of the quantum Hall
bulk in the spirit of the wire construction \citep{Teo2014}.

Throughout the paper we focus on $\nu=1$ \emph{integer} quantum Hall
regime. We investigate the predictions of above two models and show
that they predict drastically distinct energy dependences of the tunneling
behavior. We compare the predictions of these models with those of
the point-QPC model and with each other. We also provide their predictions
for a number of transport quantities that can be measured.

The structure of the paper is as follows. In Sec.~\ref{sec:transmission_function},
we remind the reader the Landauer-Büttiker formalism \citep{Nazarov2009,Buttiker1988,Martin1992}
and define the transmission function of the QPC --- the key object
for our consideration. In Sec.~\ref{sec:QPC_models}, we remind the
reader the point-QPC model and introduce our wide- and long-QPC models.
We discuss a theoretical peculiarity of the point-limit of all the
models in Sec.~\ref{sec:point-QPC-limit}. In Sec.~\ref{sec:model_comparison},
we explore the predictions of the three QPC models for realistic experiments.
Finally, we provide some concluding remarks in Sec.~\ref{sec:conclusion}.

\section{Transmission function}

\label{sec:transmission_function}
\begin{figure}[t]
\begin{centering}
\includegraphics[width=9cm,totalheight=4.5cm]{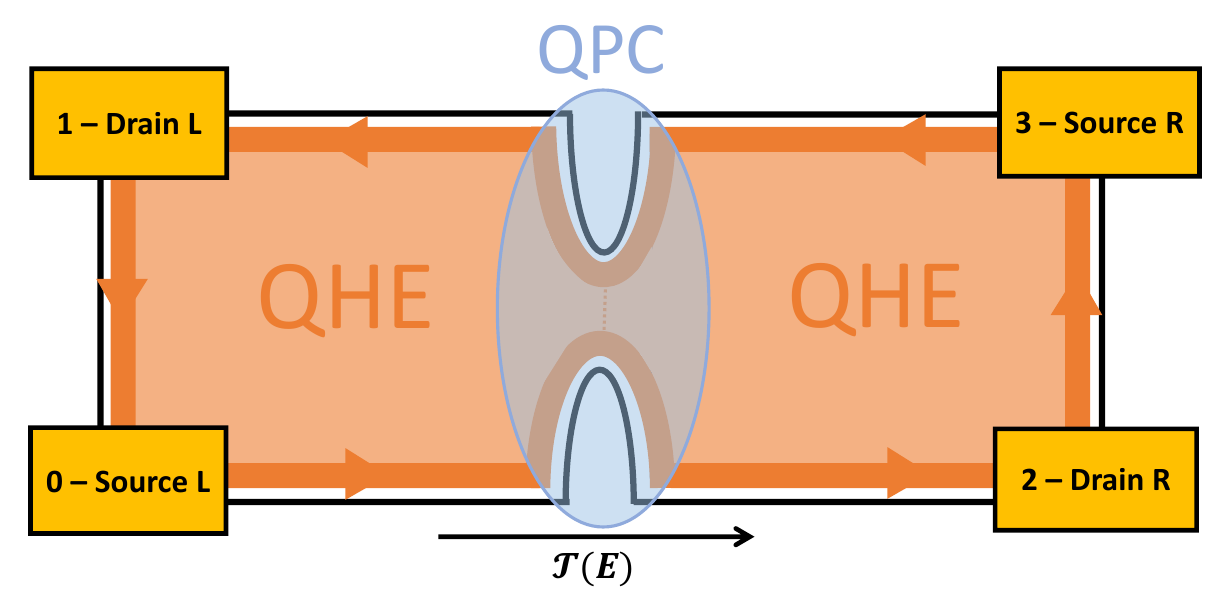}
\par\end{centering}
\caption{\label{fig:1_QPC_experiment_scheme} \textbf{Four-terminal setup of
a quantum Hall QPC experiment.} The quantum Hall bulk supports a single
chiral edge channel going around the sample (we focus on $\nu=1$
integer quantum Hall regime). The edges from the opposite sides of
the sample approach each other in one location --- this is the QPC.
Four Ohmic contacts are present in the sample, two (sources) are used
to inject current into the chiral edge, two (drains) are used to extract
the current that has passed through the QPC region. The key object
for this paper's consideration is the energy-dependent transmission
probability ${\cal T}\left(E\right)$: the probability for an excitation
to be transmitted through the QPC without jumping to the other edge
channel. }
\end{figure}

In this section, we introduce the setup of quantum Hall QPC experiments.
We focus on the case of $\nu=1$ integer quantum Hall regime and remind
the reader how the Landauer--Büttiker formalism for transport in
non-interacting mesoscopic systems applies to this system. In particular,
we define the transmission function ${\cal T}\left(E\right)$, the
key object for our consideration in the next sections.

The typical (four-terminal) setup of a quantum Hall QPC experiment
is shown in Fig.~\ref{fig:1_QPC_experiment_scheme}. The sources
are used to inject current into the chiral edge, as the drains are
used to extract the current that has passed through the QPC region.
The Landauer--Büttiker formalism \citep{Nazarov2009,Buttiker1988,Martin1992},
requires the knowledge of the scattering matrix that describes the
probability amplitudes of electrons emitted by one contact to arrive
at a different contact:
\begin{equation}
\mathcal{S}=\left(\begin{array}{cccc}
0 & \text{1} & 0 & 0\\
r_{L} & 0 & 0 & t_{R}\\
t_{L} & 0 & 0 & r_{R}\\
0 & 0 & 1 & 0
\end{array}\right)\left[\begin{array}{c}
\text{0 \textendash\ Source L}\\
\text{1 \textendash\ Drain L}\\
\text{2 \textendash\ Drain R}\\
\text{3 \textendash\ Source R}
\end{array}\right].\label{eq:S-matrix}
\end{equation}
The above matrix involves the amplitudes $t_{R/L}(E)$ for an electron
to be \emph{transmitted} through the QPC without changing its edge
and $r_{R/L}(E)$ for it to be \emph{reflected} to the other edge.
Unitarity of the S-matrix implies $\abs{t_{L}(E)}^{2}+\abs{r_{L}(E)}^{2}=1$
and $\abs{t_{R}(E)}^{2}+\abs{r_{L}(E)}^{2}=1$, from which it follows
that $\abs{t_{R}(E)}^{2}=\abs{t_{L}(E)}^{2}$. The probability of
being transmitted, ${\cal T}\left(E\right)=\abs{t_{L}(E)}^{2}=\abs{t_{R}(E)}^{2}$,
is called the transmission function.

We focus on the DC current $I_{\mathrm{2}}$ extracted from Drain
R, as well as the low-frequency noise of that current,
\begin{equation}
S_{22}=\left\langle \hat{I}(t)\hat{I}(t')+\hat{I}(t')\hat{I}(t)-2\langle\hat{I}(t)\rangle\langle\hat{I}(t')\rangle\right\rangle _{\omega},\label{eq: Noise_in_terms_of_current}
\end{equation}
where $\avg{...}_{\omega}$ is the Fourier component corresponding
to $e^{i\omega(t-t')}$ of the time-domain correlation function. The
Landauer--Büttiker formalism gives \citep{Nazarov2009}

\begin{equation}
I_{\mathrm{2}}=\frac{e}{2\pi\hbar}\int dE\ {\cal T}\left(E\right)\left(f_{\mathrm{L}}(E)-f_{\mathrm{R}}(E)\right),\label{eq:current_general}
\end{equation}

\begin{multline}
S_{\mathrm{22}}=\frac{e^{2}}{\pi\hbar}\int dE\ \bigl\{ f_{\mathrm{eq}}\left(E\right)\left(1-f_{\mathrm{eq}}\left(E\right)\right)+\\
+{\cal T}^{2}\left(E\right)f_{\mathrm{L}}\left(E\right)\left(1-f_{\mathrm{L}}\left(E\right)\right)\\
+\left(1-{\cal T}\left(E\right)\right)^{2}f_{\mathrm{R}}\left(E\right)\left(1-f_{\mathrm{R}}\left(E\right)\right)\\
+{\cal T}\left(E\right)\left(1-{\cal T}\left(E\right)\right)f_{R}\left(E\right)\left(1-f_{L}\left(E\right)\right)\\
+{\cal T}\left(E\right)\left(1-{\cal T}\left(E\right)\right)f_{L}\left(E\right)\left(1-f_{R}\left(E\right)\right)\bigr\}.\label{eq:noise_general}
\end{multline}
Here we assumed that the L and R sources are described by Fermi distributions
$f_{\mathrm{L}}\left(E\right)=f\left(E-\mu_{\mathrm{L}}\right)$ and
$f_{\mathrm{R}}\left(E\right)=f\left(E-\mu_{\mathrm{R}}\right)$,
while the drains are in equilibrium with the sample and are defined
by the equilibrium distribution $f_{\mathrm{eq}}\left(E\right)=f\left(E-\mu\right)$,
where $f\left(E\right)=\left(1+e^{E/(k_{\mathrm{B}}T)}\right)^{-1}$.
In what follows, we set $\mu=0$.

Often the approximation of constant transmission is employed: ${\cal T}\left(E\right)={\cal T}_{0}$.
In this case, the above expressions simplify to 
\begin{equation}
\left.I_{\mathrm{2}}\right|_{{\cal T}\equiv{\cal T}_{0}}=\frac{e^{2}V}{2\pi\hbar}{\cal T}_{0},\label{eq:current_constant_transmission}
\end{equation}
\begin{multline}
\left.S_{\mathrm{22}}\right|_{{\cal T}\equiv{\cal T}_{0}}=\frac{e^{2}}{\pi\hbar}\biggl[2k_{\mathrm{B}}T\\
+eV{\cal T}_{0}\left(1-{\cal T}_{0}\right)\left(\coth\left(\frac{eV}{2k_{\mathrm{B}}T}\right)-\frac{2k_{\mathrm{B}}T}{eV}\right)\biggr].\label{eq:noise_constant_transmission}
\end{multline}
Here we have taken $\mu_{\mathrm{L}}-\mu_{\mathrm{R}}=eV$ to be the
bias voltage across the QPC.

Experiments often see behavior consistent with Eqs.~(\ref{eq:current_constant_transmission}--\ref{eq:noise_constant_transmission})
for small enough $V$, whereas at larger $V$ deviations for this
behavior (or its analogue for the fractional case) are observed both
in the integer and (most prominently) in the fractional QH effect
\citep{FracChargeObs_Glattli,FracChargeObs_Heiblum,Veillon2024,Schiller2024}.
One of the possible explanations is that the dependence of ${\cal T}\left(E\right)$
on the energy becomes important at larger values of the bias voltage.
Below, we introduce several models, based on general considerations
and investigate the dependence ${\cal T}\left(E\right)$ they predict,
as well as its effect on $I_{2}$ and $S_{22}$.

\section{Analytic models of quantum point contact}

\label{sec:QPC_models}

In the previous section, we defined the key object for our consideration:
the QPC transmission $\mathcal{T}\left(E\right)$. Here we introduce
a number of phenomenological models that allow for predicting the
energy dependence of $\mathcal{T}\left(E\right)$. We start by reminding
the reader the standard point-QPC model. We then introduce two models
that account for different aspects of the realistic geometry of the
QPC: wide QPC and long QPC. We explain their physical meaning and
calculate the $\mathcal{T}\left(E\right)$ for each of the models.

\subsection{Point model}

\begin{figure}[t]
\begin{centering}
\includegraphics[width=7cm,totalheight=5cm]{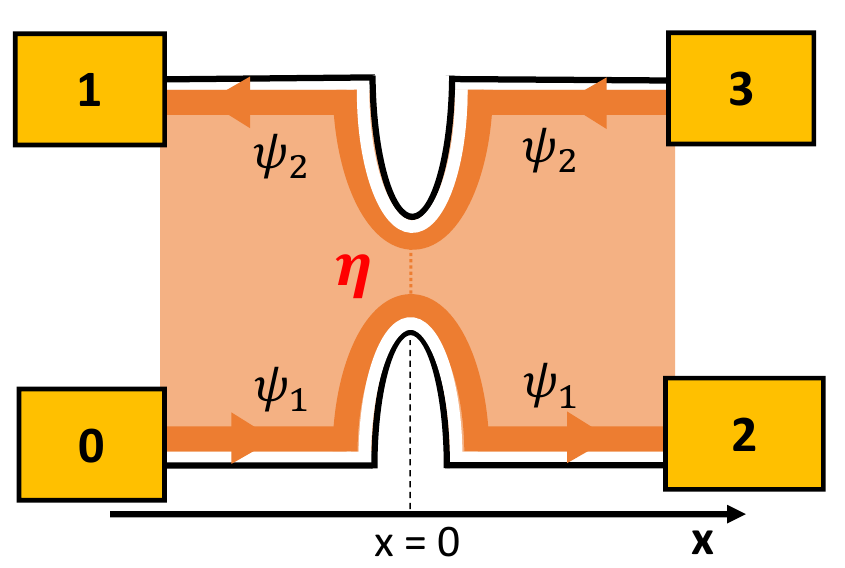}
\par\end{centering}
\centering{}\caption{\label{fig:2_point_QPC_model}\textbf{ Point-QPC model}. Two counter-propagating
chiral edge modes approach each other at a single point, $x=0$, which
allows for tunneling of excitations between the edge modes.}
\end{figure}

The standard model of QPC assumes that tunneling between the two edges
happens at a single point, cf.~\ref{fig:2_point_QPC_model}. The
model is described by the Hamiltonian 
\begin{align}
\text{H}_{\text{pqpc}} & =\text{H}_{\text{edge}}+\text{H}_{\text{tun-point}},\label{eq:hamiltonian_point_qpc}
\end{align}
\begin{gather}
\text{H}_{\text{edge}}=\int_{-\infty}^{+\infty}dx\left(-i\psi_{1}^{\dagger}v\partial_{x}\psi_{1}+i\psi_{2}^{\dagger}v\partial_{x}\psi_{2}\right),\label{eq:H_edge_pqpc}\\
\text{H}_{\text{tun-point}}=\eta\text{\ensuremath{\left(\psi_{1}^{\dagger}\left(x=0\right)\psi_{2}\left(x=0\right)+\text{h.c.}\right)}}.\label{eq:H_tun-point}
\end{gather}
The edge Hamiltonian $\text{H}_{\text{edge}}$ describes chiral fermionic
channels $\psi_{1}$ and $\psi_{2}$ propagating to the right and
to the left respectively. The tunneling Hamiltonian $\text{H}_{\text{tun-point}}$
describes electron transport between the two edges. Solving the scattering
problem for $\text{H}_{\text{pqpc}}$, one finds the transmission
function
\begin{equation}
\text{\ensuremath{\mathcal{T}_{\text{pqpc}}\left(E\right)}}=\mathcal{T}_{\text{pqpc}}^{0\rightarrowtriangle2}=\left(\frac{1-\frac{\eta{}^{2}}{4v^{2}}}{1+\frac{\eta{}^{2}}{4v^{2}}}\right)^{2}.\label{eq:transmission_point_qpc}
\end{equation}
The reader can find the details of the derivation in Appendix~\ref{sec:transmission_pqpc_derivation}.

The above result, though standard, warrants a few comments. First,
note that $\text{\ensuremath{\mathcal{T}_{\text{pqpc}}\left(E\right)}}$
does not depend on energy, which renders the transport predictions
(\ref{eq:current_constant_transmission}, \ref{eq:noise_constant_transmission})
exact for this model.

Second, one naturally expects bigger values of $\eta$ to lead to
bigger tunneling (and, thus, smaller transmission) at the QPC. This
intuition holds for $\abs{\eta}\leq2v$, at which $\ensuremath{\mathcal{T}_{\text{pqpc}}\left(E\right)=0}$
vanishes. For $\abs{\eta}>2v$, however, one again has $\mathcal{T}_{\text{pqpc}}\left(E\right)>0$.
This indicates singular behavior of this model, which has been previously
pointed out in the context of perturbative calculations \citep{Filippone2016}.
The wide-QPC and long-QPC models considered below regularize this
behavior \emph{in distinct ways}. We will discuss this in more detail
in Sec.~\ref{sec:model_comparison}.

\subsection{Wide model}

\begin{figure}[t]
\begin{centering}
\includegraphics[width=7cm,totalheight=6cm]{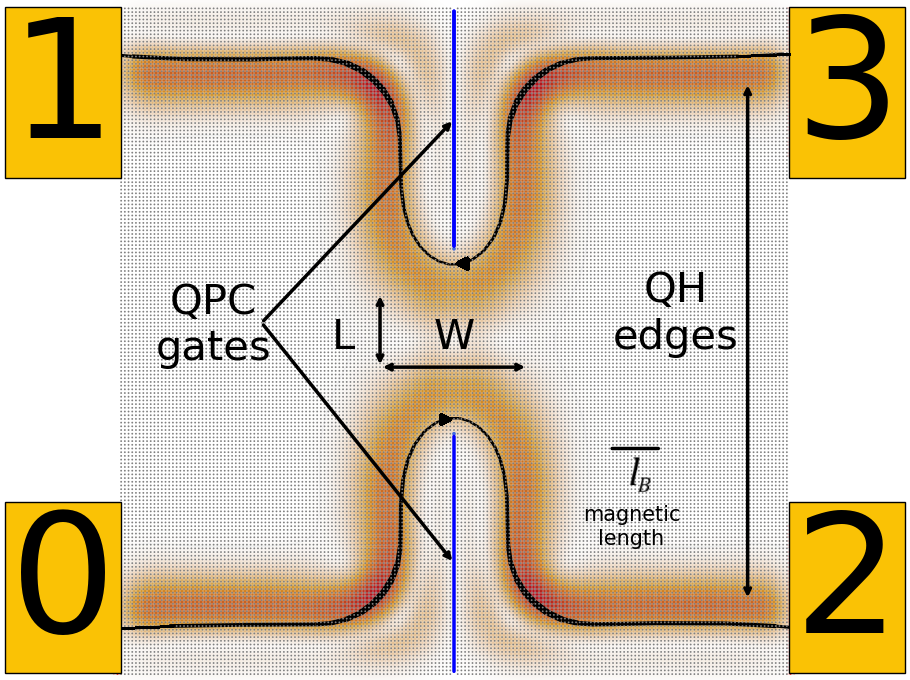}
\par\end{centering}
\begin{centering}
\caption{\label{fig:edge_modes}\textbf{Numerical simulation of a QPC geometry
with narrow QPC-defining gates}. The figure presents the current density
of QH edges in a sample that has four leads and a QPC defined by infinite
potential on narrow gates (blue lines). The sample is modeled as 180x180
lattice, and the QPC-defining potential is applied to regions of 1-site
width and 70-site length. The red density map represents the current
density of two edge states. The edge states go around the gated regions
smoothly, so that the tunneling would happen over a finite QPC width
of a few magnetic lengths ($W$). This motivates our wide-QPC model
shown in Fig.~\ref{fig:wide-QPC_model}. The distance $L$ between
the edges in the QPC determines the size of the gapped system bulk
between the edges. This motivated our long-QPC model shown in Fig.~\ref{fig:long-QPC_model}.
The simulation has been performed using KWANT package \citep{Groth2014}.}
\par\end{centering}
\end{figure}

In the previous section, we presented the conventional point-contact
(point-QPC) model. A key assumption of the point-QPC model is that
tunneling occurs at a single spatial point. However, quantum point
contacts in real samples inevitably possess a finite tip width. Moreover,
even in the most narrowly constructed QPCs, opposite QH edges are
in contact over a width of the order of the magnetic length $\ensuremath{l_{B}=\sqrt{\hbar/(eB)}}$,
as exemplified in Fig. \ref{fig:edge_modes}. This challenges the
point-like approximation and warrants an investigation of the effects
of the finite width of the QPC.

To this end, we introduce the wide-QPC model, which incorporates a
finite-width tunneling region between the chiral QH edge channels(Fig.
\ref{fig:wide-QPC_model}). In this model, the two edge states, $\psi_{1}$
and $\psi_{2}$ propagate in opposite directions, as in Fig. \ref{fig:2_point_QPC_model}.
However, the tunneling happens over edge length L, and the tunneling
amplitude is $\zeta$: 
\begin{figure}[t]
\begin{centering}
\includegraphics[width=7cm,totalheight=5cm]{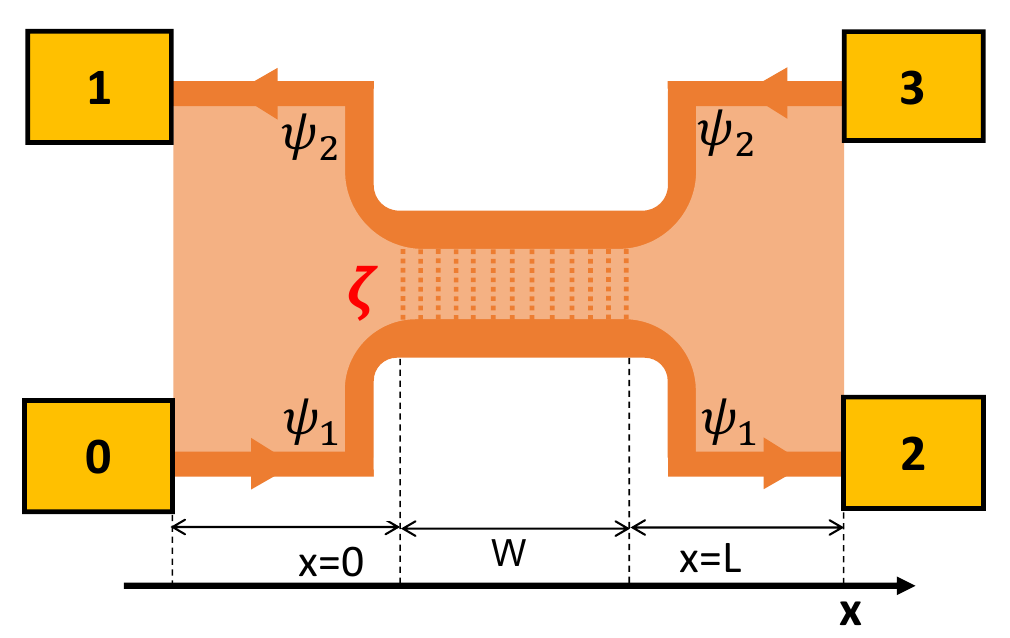}
\par\end{centering}
\centering{}\caption{\label{fig:wide-QPC_model} \textbf{Wide-QPC model.} Two chiral edge
channels undergo tunneling over a finite region of width $W$.}
\end{figure}

\begin{equation}
\text{H}_{\text{wqpc}}=\text{H}_{\text{edge}}+\text{H}_{\text{tun-wide}},\label{eq:hamiltonian_wide_qpc}
\end{equation}
\begin{gather}
\text{\text{H}}_{\text{edge}}=\int_{-\infty}^{+\infty}dx\left(-i\psi_{1}^{\dagger}v\partial_{x}\psi_{1}+i\psi_{2}^{\dagger}v\partial_{x}\psi_{2}\right),\label{eq:H_edge_wqpc}\\
\text{H}_{\text{tun-wide}}=\zeta\int_{0}^{W}dx\left(\psi_{1}^{\dagger}\left(x\right)\psi_{2}\left(x\right)+\text{h.c.}\right).\label{eq:H_tun_wide}
\end{gather}
Solving the scattering problem for $\text{H}_{\text{wqpc}}$ (see
Appendix~\ref{sec:transmission_wqpc_derivation} for derivation),
one finds the transmission function for wide-QPC model $\text{\ensuremath{\left(\text{\ensuremath{\mathcal{T}}}_{\text{0\ensuremath{\rightarrowtriangle}2}}^{\text{wqpc}}\right)}}$
as:
\begin{align}
\mathcal{T}_{\mathrm{wqpc}}\left(E\right)=\mathcal{T}_{\text{wqpc}}^{0\rightarrowtriangle2}= & \frac{1}{1+\Xi(E,\zeta,W)},\label{eq:transmission_wide_qpc}\\
\text{where }\Xi(E,\zeta,W) & =\frac{\sinh^{2}\left(\frac{\zeta W}{v}\sqrt{1-\left(\frac{E}{\zeta}\right)^{2}}\right)}{\left(1-\left(\frac{E}{\zeta}\right)^{2}\right)}\nonumber \\
 & =\frac{\sin^{2}\left(\frac{\zeta W}{v}\sqrt{\left(\frac{E}{\zeta}\right)^{2}-1}\right)}{\left(\left(\frac{E}{\zeta}\right)^{2}-1\right)}.\nonumber 
\end{align}
Note that unlike in the case of point-QPC model, here the transmission
depends on energy.

The typical scale for the energy is the tunneling amplitude $\zeta$,
whereas the overall behavior of the transmission is determined by
the dimensionless parameter $\zeta W/v$. Indeed, at $\mathcal{T}_{\mathrm{wqpc}}\left(E=0\right)=\left[1+\sinh^{2}\left(\frac{\zeta W}{v}\right)\right]^{-1}$.
At large energies, the transmission is perfect: $\mathcal{T}_{\mathrm{wqpc}}\left(\abs E\gg\zeta\right)=1$.
At the characteristic scale, $\mathcal{T}_{\mathrm{wqpc}}\left(\abs E=\zeta\right)=\left[1+\left(\frac{\zeta W}{v}\right)^{2}\right]^{-1}$.

\begin{figure}[t]
\begin{centering}
\includegraphics[width=8cm,totalheight=7cm]{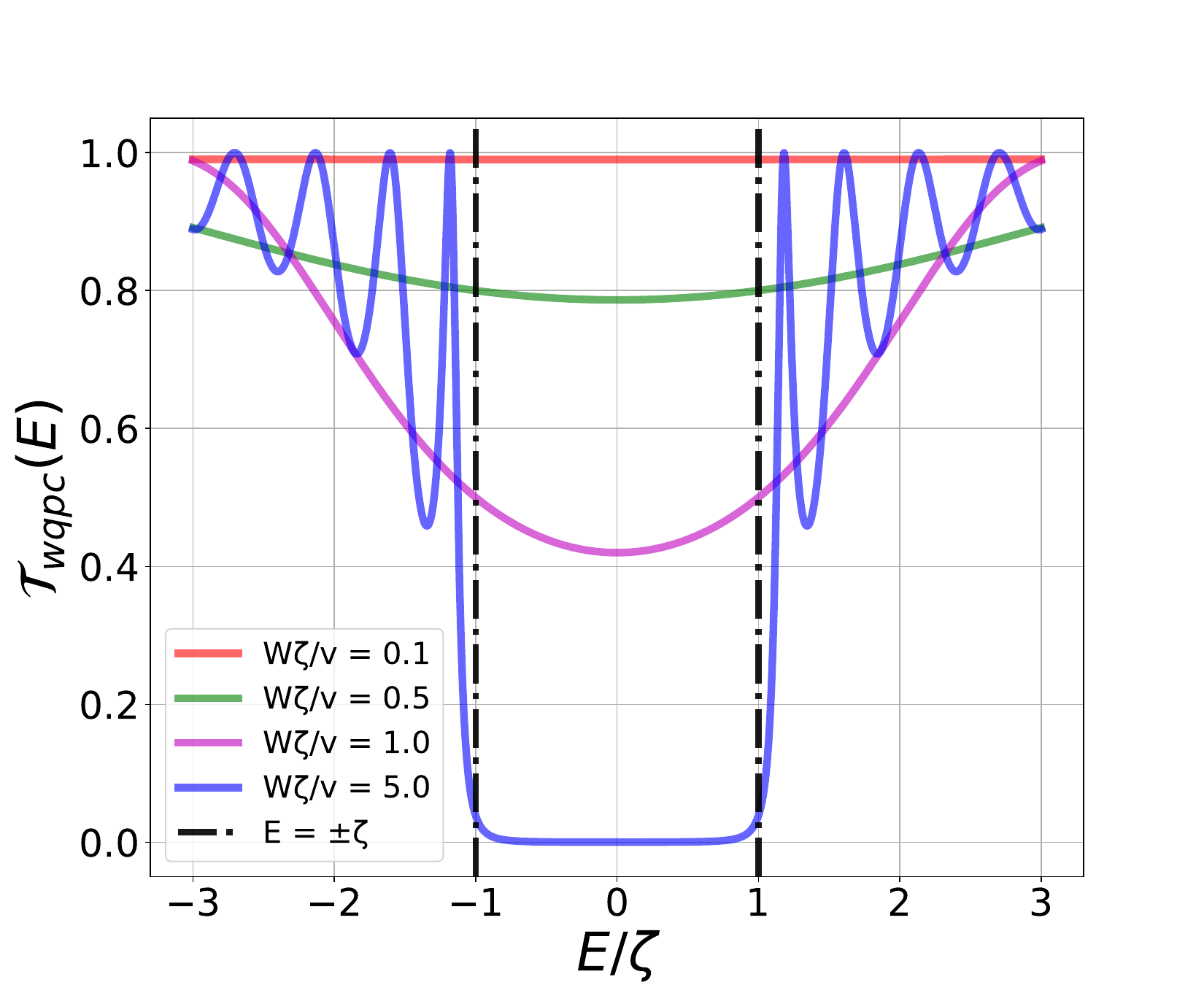}
\par\end{centering}
\centering{}\caption{\label{fig:transmission_wide_qpc}\textbf{The dependence of $\mathcal{T}_{\mathrm{wqpc}}\left(E\right)$
on energy for various values of $\zeta W/v$.} The value of the dimensionless
parameter $\zeta W/v$ determines $\mathcal{T}_{\mathrm{wqpc}}\left(E=0\right)$.
At $\protect\abs E\rightarrow\infty$, the transmission becomes perfect,
$\mathcal{T}_{\mathrm{wqpc}}\left(E\right)\rightarrow1$. Oscillatory
behavior for $\protect\abs E>\zeta$ strongly depends on $\zeta W/v$.}
\end{figure}

The behavior of $\mathcal{T}_{\mathrm{wqpc}}\left(E\right)$ at various
values of $\zeta W/v$ is shown in Fig.~\ref{fig:transmission_wide_qpc}.
Several aspects are worth noting. The strength of energy dependence
of $\mathcal{T}_{\mathrm{wqpc}}\left(E\right)$ varies with $\zeta W/v$
in a non-trivial way. In particular, for $\abs E>\zeta$, $\mathcal{T}_{\mathrm{wqpc}}\left(E\right)$
oscillates, cf.~Eq.~(\ref{eq:transmission_wide_qpc}). The oscillation
frequency, however, depends on $\zeta W/v$ and is only noticeable
for large $\zeta W/v$ in Fig.~\ref{fig:transmission_wide_qpc}.

\subsection{Long model}

\begin{figure}[t]
\begin{centering}
\includegraphics[width=6cm,totalheight=4cm]{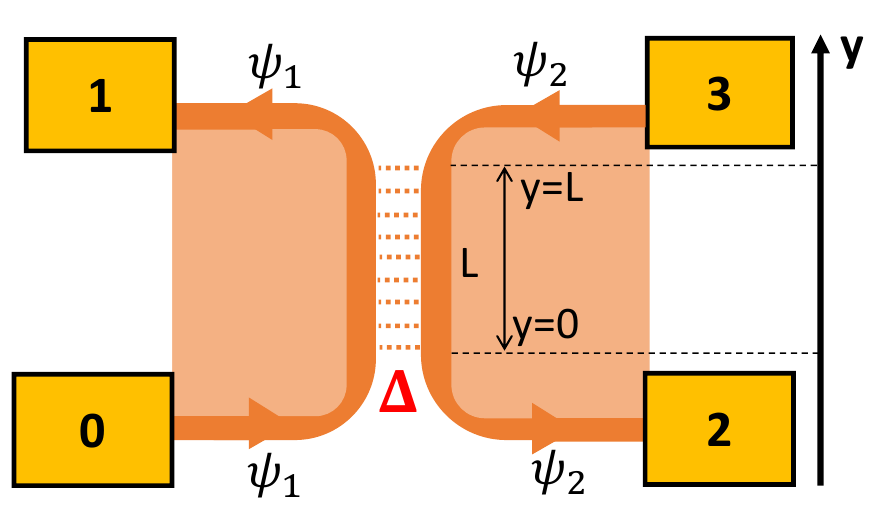}
\par\end{centering}
\centering{}\caption{\label{fig:long-QPC_model}\textbf{Long QPC model.} In order to model
the tunneling barrier created by the bulk gap inside the QPC (cf.~Fig.~\ref{fig:1_QPC_experiment_scheme}),
we model the bulk in the spirit of the wire construction. The region
of strong electron tunneling (amplitude $\Delta$) across the region
devoid of electrons is topologically equivalent to the bulk of the
quantum Hall sample. }
\end{figure}

In the models considered above (point QPC and wide QPC), the tunneling
amplitude between the edges was treated as a phenomenological parameter.
However, the intuition based on quantum mechanics tells that the tunneling
arises from a barrier that makes it energetically costly for quasiparticles
to be between the two edges in the QPC. This barrier may govern the
energy dependence of the tunneling amplitude, and thus of the transmission
probability ${\cal T}\left(E\right)$.

In order to model this dependence, we model the QH bulk region of
size $L$ inside the QPC, cf.~Fig.~\ref{fig:edge_modes}, in the
spirit of the wire construction \citep{Teo2014}. In its general form,
the wire construction models the QH bulk as a 1D array of one-dimensional
wires with appropriate tunneling processes between them. Here we use
a minimalistic model, framing the QPC as two edges, between which
electrons tunnel across a depleted region, cf.~Fig.~\ref{fig:long-QPC_model}.
The tunneling amplitude $\Delta$ corresponds to the bulk gap, cf.~Appendix~\ref{sec:transmission_wqpc_tunnelling_region_properties},
and the length of the tunneling region $L$ corresponds to the distance
between the edges on the different sides of the QPC --- these parameters
have notably been absent from the point- and wide-QPC models.\footnote{While here we consider the example of integer QH effect, the generalization
of this model for the fractional QH effect would reflect the famous
duality of weak quasiparticle tunneling and strong electron tunneling
\citep{Wen2004,Saleur,Saleur2,Saleur_LaughlinNoise_detailed}.}

The Hamiltonian for the long-QPC model is
\begin{equation}
\text{H}_{\text{lqpc}}=\text{H}_{\text{edge}}+\text{H}_{\text{tun-long}},\label{eq:hamiltonian_long_qpc}
\end{equation}
\begin{gather}
\text{\text{H}}_{\text{edge}}=\int_{-\infty}^{+\infty}dy\left(-i\psi_{1}^{\dagger}v\partial_{y}\psi_{1}+i\psi_{2}^{\dagger}v\partial_{y}\psi_{2}\right),\label{eq:H_edge_lqpc}\\
\text{H}_{\text{tun-long}}=\Delta\int_{0}^{L}dy\left(\psi_{1}^{\dagger}\left(y\right)\psi_{2}\left(y\right)+\text{h.c.}\right).\label{eq:H_tun_long}
\end{gather}

Note that the Hamiltonian for the long-QPC model is formally identical
to that of the wide-QPC model, Eqs.~(\ref{eq:hamiltonian_wide_qpc}--\ref{eq:H_tun_wide}).
The two Hamiltonians differ by the substitutions $x\rightarrow y$,
$\zeta\rightarrow\Delta$, and $W\rightarrow L$, which reflects the
conceptual difference between the models. Most importantly, the models
differ by the connection of $\psi_{1}$ and $\psi_{2}$ to different
Ohmic contacts, cf.~Figs.~\ref{fig:wide-QPC_model} and \ref{fig:long-QPC_model}.

This enables us to reuse the solution of the wide-QPC model (\ref{eq:transmission_wide_qpc}),
while accounting for these distinctions: 
\begin{multline}
\mathcal{T}_{\mathrm{lqpc}}\left(E\right)=\mathcal{T}_{\text{lqpc}}^{0\rightarrowtriangle2}=1-\mathcal{T}_{\text{lqpc}}^{0\rightarrowtriangle1}\\
=1-\left.\mathcal{T}_{\text{wqpc}}^{0\rightarrowtriangle2}\right|_{\begin{array}{c}
\zeta\rightarrow\Delta\\
W\rightarrow L
\end{array}}=\frac{\Xi(E,\Delta,L)}{1+\Xi(E,\Delta,L)},\label{eq:transmission_long_qpc}
\end{multline}
\begin{align*}
\text{where }\Xi(E,\Delta,L) & =\frac{\sinh^{2}\left(\frac{L\Delta}{v}\sqrt{1-\left(\frac{E}{\Delta}\right)^{2}}\right)}{\left(1-\left(\frac{E}{\Delta}\right)^{2}\right)}\\
 & =\frac{\sin^{2}\left(\frac{L\Delta}{v}\sqrt{\left(\frac{E}{\Delta}\right)^{2}-1}\right)}{\left(\left(\frac{E}{\Delta}\right)^{2}-1\right)}.
\end{align*}
Similarly to the wide-QPC model and in contrast to the standard point-QPC
model, the transmission function in the long-QPC model explicitly
depends on energy. However, the nature of the dependence is drastically
different as compared to the wide-QPC model.

\begin{figure}[t]
\begin{centering}
\includegraphics[width=8cm,totalheight=7cm]{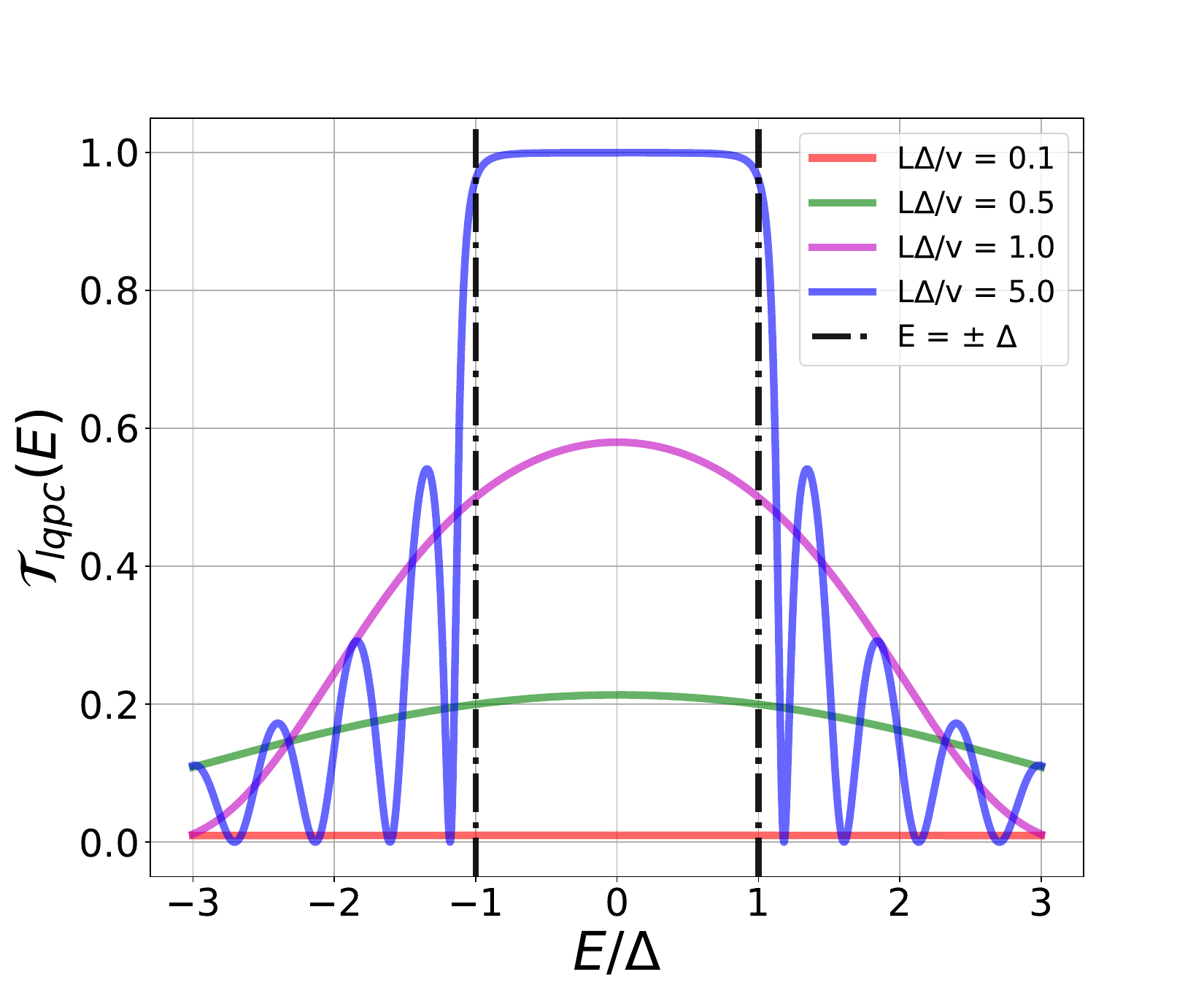}
\par\end{centering}
\centering{}\caption{\textbf{\label{fig:transmission_long_qpc}The dependence of $\mathcal{T}_{\mathrm{lqpc}}\left(E\right)$
on energy for various values of $L\Delta/v$.} The dimensionless parameter
$L\Delta/v$ is associated to the \textquotedblleft area of the tunneling
barrier\textquotedblright , $L\Delta$. Its value determines $\mathcal{T}_{\mathrm{lqpc}}\left(E=0\right)$.
At $\protect\abs E\rightarrow\infty$, the transmission decays to
zero, $\mathcal{T}_{\mathrm{lqpc}}\left(E\right)\rightarrow0$. The
vertical black lines $E=\pm\Delta$, marks the magnitude of the bulk
gap $\Delta$; we do not expect our model to be valid beyond this
energy scale.}
\end{figure}

Indeed, the transmission is \emph{maximum} at $E=0$ and goes to zero
at large energies, $\mathcal{T}_{\mathrm{lqpc}}\left(\abs E\gg\Delta\right)=0$,
cf.~Fig.~\ref{fig:transmission_long_qpc}. This is natural as for
$\abs E\ll\Delta$ the quasiparticles are unlikely to tunnel across
the QPC, and are likely to be transmitted. Conversely, for $\abs E\gg\Delta$,
the bulk gap does not present a significant obstacle for quasiparticles
to travel to a different edge. Of course, our crude model does not
capture all the important aspects of the QH bulk, notably its disordered
nature. However, our model does capture the concept of the tunneling
barrier.

The value of transmission $\mathcal{T}_{\mathrm{lqpc}}\left(E=0\right)=\left[1+\sinh^{-2}\left(\frac{L\Delta}{v}\right)\right]^{-1}$
is controlled by the dimensionless parameter $L\Delta/v$ that captures
the barrier height $\Delta$ and length $L$. We do not expect our
model to be valid beyond $\abs E=\Delta$, where proper bulk modeling
needs to be performed. At this energy scale, $\mathcal{T}_{\mathrm{lqpc}}\left(\abs E=\Delta\right)=\left[1+\left(\frac{L\Delta}{v}\right)^{-2}\right]^{-1}$.

\section{Point-QPC limit}

\label{sec:point-QPC-limit}

The models introduced in the previous section can be viewed as generalizations
of the point-QPC model, incorporating aspects of realistic QPC setups.
The wide- and long-QPC models allow a parameter regime ($L\rightarrow0$)
that reduces them back to the point-QPC geometry. It is natural to
expect the predictions for ${\cal T}\left(E\right)$ to reduce to
those of the point-QPC model in this limit. Below we show that this
is not the case, which points at the singularity of the point-QPC
model. 

Consider the wide-QPC model. Taking the limit $W\rightarrow0$, $\zeta\rightarrow\infty$
such that $\zeta W=\mathrm{const}$, the Hamiltonian of the wide-QPC
model (\ref{eq:hamiltonian_wide_qpc}--\ref{eq:H_tun_wide}) reduces
to that of the point-QPC model (\ref{eq:hamiltonian_point_qpc}--\ref{eq:H_tun-point})
with $\eta_{\text{wqpc}}=\zeta W$. Taking the limit for the transmission
function (\ref{eq:transmission_wide_qpc}) yields
\begin{equation}
\mathcal{T}_{\mathrm{wqpc}}\left(E\right)=\mathcal{T}_{\mathrm{wqpc}}^{(\text{point limit})}=\frac{1}{1+\sinh^{2}\left(\frac{\eta_{\text{wqpc}}}{v}\right)}.
\end{equation}

Compare this to $\text{\ensuremath{\mathcal{T}_{\text{pqpc}}\left(E\right)}}$
in Eq.~(\ref{eq:transmission_point_qpc}). Both expressions predict
no energy dependence of the transmission function and both predict
$\mathcal{T}\left(E\right)\approx1-\eta^{2}/v^{2}$ for $\abs{\eta}\ll v$.
However, for $\eta$ comparable or larger than $v$, the models yield
drastically different predictions, despite having the identical Hamiltonian
in this limit. In particular, the point-QPC model predicts $\text{\ensuremath{\mathcal{T}_{\text{pqpc}}\left(E\right)}}=0$
at $\eta=2v$ and $\text{\ensuremath{\mathcal{T}_{\text{pqpc}}\left(E\right)}}=1$
at $\eta\rightarrow\infty$. At the same time, $\mathcal{T}_{\mathrm{wqpc}}^{(\text{point limit})}$
is monotonous with respect to $\eta_{\text{wqpc}}$ and vanishes for
$\eta_{\text{wqpc}}\rightarrow\infty$.

The long-QPC model also features a point limit $L\rightarrow0$, $\Delta\rightarrow\infty$
such that $L\Delta=\mathrm{const}$. In this limit, its Hamiltonian
(\ref{eq:hamiltonian_long_qpc}--\ref{eq:H_tun_long}) reduces to
that of the point-QPC model (\ref{eq:hamiltonian_point_qpc}--\ref{eq:H_tun-point})
with $\eta_{\text{long}}=L\Delta$. However, the roles of the contacts
1 and 2 are interchanged compared to the wide- and point-QPC models,
cf.~Figs.~\ref{fig:2_point_QPC_model}, \ref{fig:wide-QPC_model},
\ref{fig:long-QPC_model}. Taking the limit for the transmission function,
one gets
\begin{equation}
\mathcal{T}_{\mathrm{lqpc}}\left(E\right)=\mathcal{T}_{\mathrm{lqpc}}^{(\text{point limit})}=\frac{\sinh^{2}\left(\frac{\eta_{\text{lqpc}}}{v}\right)}{1+\sinh^{2}\left(\frac{\eta_{\text{lqpc}}}{v}\right)}.
\end{equation}
Again, this predicts an energy-independent transmission function in
the point limit, yet does not match $\mathcal{T}_{\text{pqpc}}^{0\rightarrowtriangle1}=1-\mathcal{T}_{\text{pqpc}}^{0\rightarrowtriangle2}$
in its dependence on $\eta$, see Eq.~(\ref{eq:transmission_point_qpc}).

This shows that the point-QPC model (\ref{eq:hamiltonian_point_qpc}--\ref{eq:H_tun-point})
is ill-defined: depending on the regularization of the point tunneling,
the model produces different results. This has, in fact, been highlighted
before in Ref.~\citep{Filippone2016}.

These results are in stark contrast to the case of point barrier in
the conventional quantum mechanics for $p^{2}/(2m)$ dispersion relation,
where the transmission across the delta-barrier is reproduced by taking
the point limit of a rectangular potential, see Appendix~\ref{sec:point_limit_p^2}.
It may be interesting to investigate in the future whether incorporating
the curvature of the dispersion relation into the model (i.e., the
terms $\sim\psi_{i}^{\dagger}\partial_{x}^{2}\psi_{i}$) would eliminate
the above singularity of the point-QPC model.

\section{Model comparison with realistic parameters}

\label{sec:model_comparison}

Above, we have defined three distinct QPC models and derived their
predictions for the transmission function ${\cal T}\left(E\right)$.
In this section, we compare these models using experimentally realistic
parameters. In Sec.~\ref{sec:transmission_comparison}, we compare
their predictions for the transmission functions. In Sec.~\ref{sec:transport_predictions},
we show how these predictions translate to experimentally observable
quantitites.

\subsection{Model comparison with experimentally realistic parameters}

\label{sec:transmission_comparison}

Above, we have derived the energy dependence of the transmission function
$\mathcal{T}\left(E\right)$ as predicted by the point-QPC, wide-QPC,
and long-QPC models. These predictions exhibit qualitatively different
behaviors: for the wide-QPC model, the minimum value of $\mathcal{T}\left(E\right)$
is achieved at $E=0$ (Fig~\ref{fig:transmission_wide_qpc}) whereas
in long-QPC model $E=0$ corresponds to the maximum of $\mathcal{T}\left(E\right)$
(Fig~\ref{fig:transmission_long_qpc}). Therefore, different aspects
of realistic QPC geometry predict different energy dependences of
$\mathcal{T}\left(E\right)$. However, what is the significance of
those dependences in realistic systems? We investigate this question
below.

We begin by estimating the experimental parameters appropriate for
typical QPC setups. Consider a GaAs-based quantum Hall sample with
an electron density $n_{s}=1\times10^{15}\,\text{m}^{-2}$. Given
the effective mass $m^{*}=0.067m_{e}$ (with $m_{e}$ is the bare
electron mass), the filling factor $\nu=1$ is achieved for the magnetic
field $B=4.14\,\text{T}$, which further sets the magnetic length
$l_{B}=12.6$~nm and the cyclotron energy $E_{c}=\hbar\omega_{c}=7.17$~meV. 

The edge state velocity may vary depending on the sample: values of
the orders $v=10^{4}$--$10^{6}$~m/s have been reported depending
on whether the quantum Hall edge is defined by a gate or by the boundary
of the sample \citep{McClure2009,Kamata2010,Kumada2011,Kataoka2016,Gurman2016,Nakamura2019,Nakamura2020,Weldeyesus2025}.
Given that the velocity of interest for us is the velocity inside
the QPC that is typically gate-defined, we take the value on the lower
end of the range, $v=2\cdot10^{4}$~m/s \citep{Gurman2016}.

Given the system parameters, we can now estimate the relevant parameters
of our QPC models. For the wide-QPC model, the parameters are the
tunneling amplitude, $\zeta$, and the width of the tunneling region,
$W$. The QPC width can hardly be smaller than a few magnetic lengths.
Indeed, Fig.~\ref{fig:edge_modes} shows that even for infinitely-small
QPC-defining gate, the tunneling region has double the width of the
edge wave function. With the edge wave function having the width $l_{B}\sqrt{2}$
\citep[Section 9.5.4]{Band2013}, we take $W=2\sqrt{2}l_{B}$. The
tunneling amplitude $\zeta$ is then determined by the transmission
value at zero energy, $\mathcal{T}_{0}=\mathcal{T}\left(E=0\right)$,
from Eq.~(\ref{eq:transmission_wide_qpc}):

\begin{equation}
\zeta=\frac{\hbar v}{W}\sinh^{-1}\left(\sqrt{\frac{1-\mathcal{T}_{0}}{\mathcal{T}_{0}}}\right),
\end{equation}
 where we have explicitly restored $\hbar$. This gives the values
of $\zeta=0.45$, $0.22$, and $0.08$~meV for $\mathcal{T}_{0}=0.1$,
$\mathcal{T}_{0}=0.5$, and $\mathcal{T}_{0}=0.9$ respectively.

For the long-QPC model, the parameters are the tunneling amplitude,
$\Delta$, and the QPC length, $L$. The value of $2\Delta$ is natural
to identify with the bulk gap of the quantum Hall sample, cf.~Appendix~\ref{sec:transmission_wqpc_tunnelling_region_properties}.
Therefore, $\Delta=E_{c}/2=3.58$~meV. The length of the QPC is then
determined by the zero-energy transmission $\mathcal{T}_{0}$ through
Eq.~(\ref{eq:transmission_long_qpc}):

\begin{equation}
L=\frac{\hbar v}{\Delta}\sinh^{-1}\left(\sqrt{\frac{\mathcal{T}_{0}}{1-\mathcal{T}_{0}}}\right).
\end{equation}
This yields the values of $L=1.2$, $\text{3.2}$, and $6.7$~nm
for $\mathcal{T}_{0}=0.1$, $\mathcal{T}_{0}=0.5$, and $\mathcal{T}_{0}=0.9$
respectively.

\begin{figure}[t]
\begin{centering}
\includegraphics[width=9cm,totalheight=6cm]{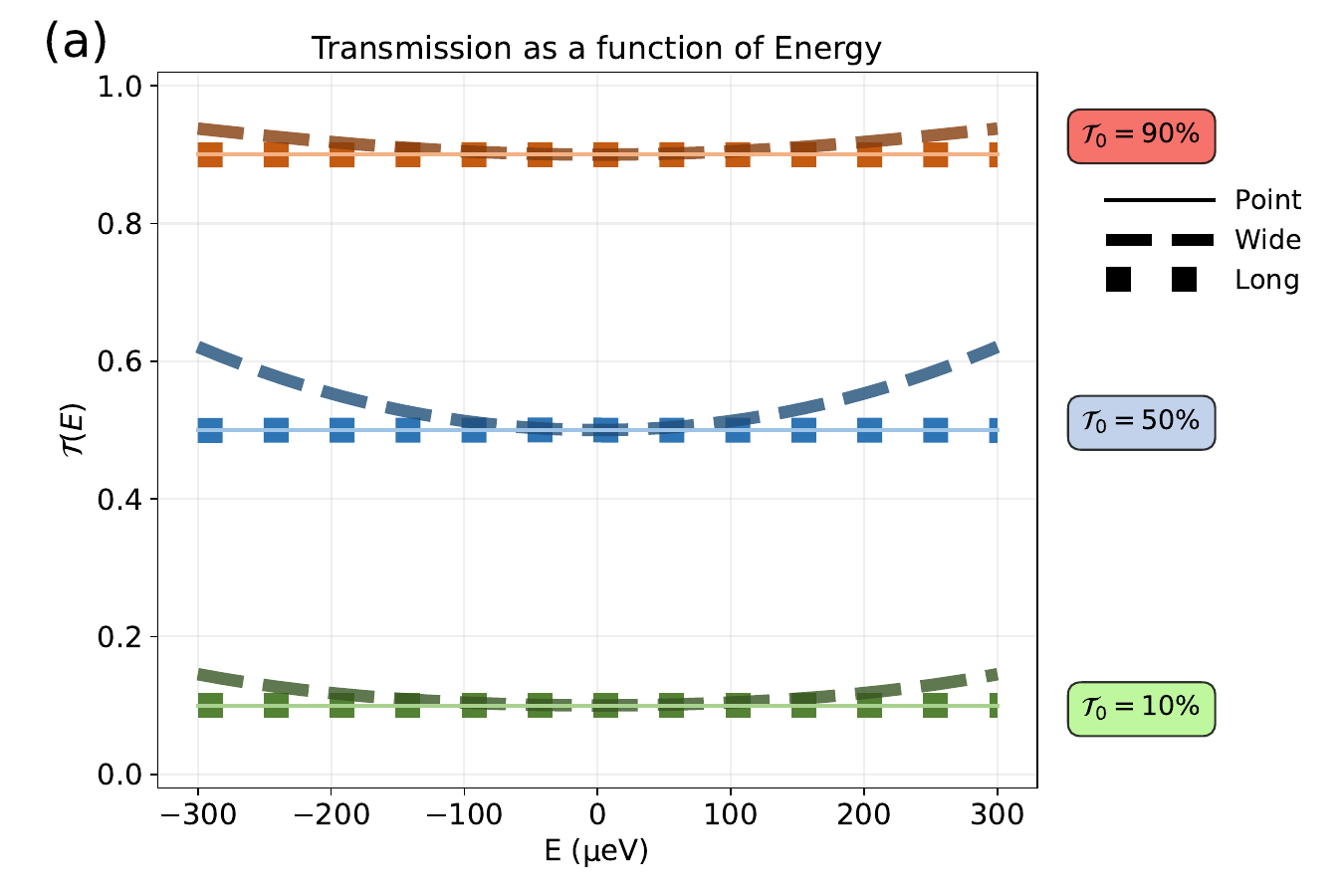}
\par\end{centering}
\centering{}\includegraphics[width=9cm,totalheight=6cm]{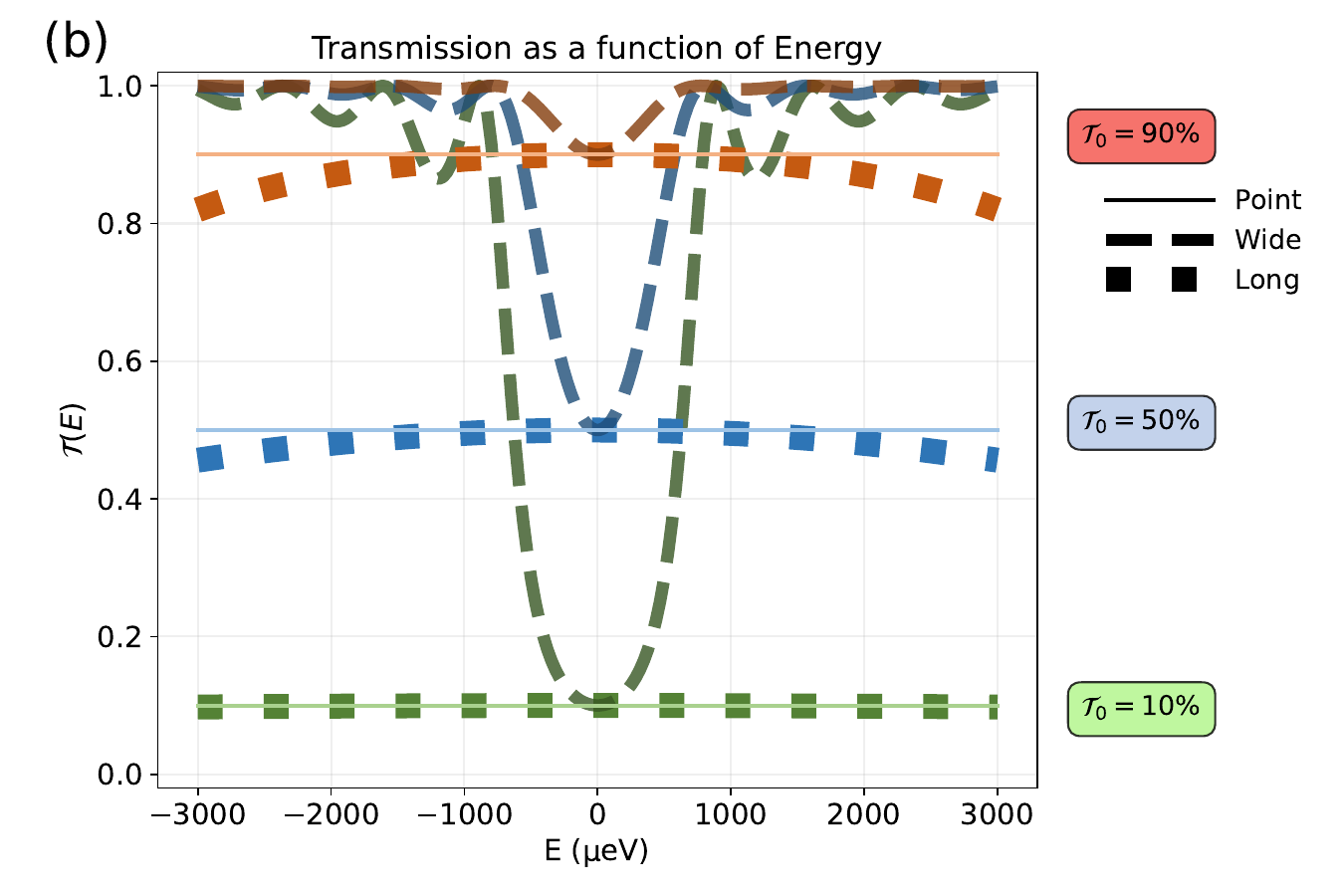}\caption{\label{fig:T(E)_realistic_parameters} \textbf{Transmission $\mathcal{T}(E)$
for various models for experimentally realistic parameters.} (a) Comparison
of the dependence of QPC transmission on energy that point-, wide-,
and long-QPC models predict for realistic experimental parameters.
The effects of the long-QPC model are negligible in the experimentally
relevant regime. Whereas the width of the QPC does have a noticeable
effect. These predictions are obtained using edge velocity $v=2\cdot10^{4}$~m/s.
The role of velocity is further discussed in Appendix~\ref{sec:role_of_edge_velocity}.
(b) Comparison of QPC energy-dependent transmission for the same set
of models for energy scales beyond the typical experimental range.
The effects of the long-QPC model become noticeable as the scale becomes
comparable with the cyclotron gap $E_{c}=\hbar\omega_{c}=7.17\,\text{meV}$.
The wide-QPC model predicts an even stronger dependence with transmission
saturating at 1 (up to small oscillations) at large energies.}
\end{figure}

Given the parameters estimated above, the dependences $\mathcal{T}\left(E\right)$
are shown in Fig.~\ref{fig:T(E)_realistic_parameters}(a). The range
of energies on the horizontal axis is chosen to correspond to the
typical bias voltages (100--300~$\mu$V) used in experiments. One
sees that the long-QPC does not predict any noticeable deviation of
$\mathcal{T}\left(E\right)$ from constant. This is due to the energies
of interest being much smaller than the bulk gap $2\Delta=E_{c}$.
The wide-QPC model, though, produces noticeable deviations. This shows
that the QPC width is an important effect even for the narrowest QPCs
possible.

We investigate the variation of $\mathcal{T}\left(E\right)$ at energies
beyond the typical voltage range in experiments in Fig.~\ref{fig:T(E)_realistic_parameters}(b).
One sees a much stronger variation of transmission. Here, the long-QPC
model predicts a noticeable deviation from the point-QPC model. The
wide-QPC model exhibits an extremely strong variation of transmission,
with saturation at $\mathcal{T}\left(E\right)=1$ and oscillations
indicative of resonances in the QPC region.

These $\mathcal{T}\left(E\right)$ dependences translate into experimentally
measurable quantities: the QPC current, the current noise, and the
Fano factor. We investigate predictions for those quantities in Sec.~\ref{sec:transport_predictions}.

\subsection{Transport predictions }

\label{sec:transport_predictions}

In the previous section, we analyzed the transmission functions associated
with three distinct models with realistic experimental parameters
in the conventional experimental regime. In this section, we investigate
the predictions for transport quantities --- specifically, the differential
conductance, current, noise, and the Fano factor --- within the point-,
wide-, and long-QPC models. We use the same parameters as above for
the three models, and take the chemical potential values for the contacts
as $\mu_{L}=-\mu_{R}=eV/2$, where $V$ is the bias voltage. We take
the system temperature to be $T=10$~mK (experiments typically use
$T=10$--$40$~mK).

Using Eqs.~(\ref{eq:current_general}--\ref{eq:noise_general}),
the current $I_{2}(V)$ and noise $S_{22}$ for each model are obtained
directly from the corresponding transmission function $\mathcal{T}\left(E\right)$.
The differential conductance is defined as 
\begin{equation}
g=dI_{2}/dV.
\end{equation}
We define the excess noise as the component of the noise beyond the
Johnson--Nyquist contribution:
\[
S_{E}=S_{22}\left(V\right)-S_{22}\left(V=0\right).
\]
Further, we consider a dimensionless parameter called the Fano factor.
We define it as,
\begin{equation}
F=\frac{S_{E}\left(V\right)}{2e\left(\frac{e^{2}}{h}V\right)\bar{\mathcal{T}}\left(V\right)\left(1-\bar{\mathcal{T}}\left(V\right)\right)};\quad\bar{\mathcal{T}}(V)=\frac{I_{2}}{I_{0}}\label{eq: Fano_factor}
\end{equation}
where $\bar{\mathcal{T}}\left(V\right)$ is average transmission at
a given voltage, defined as ratio of the transmitted current $I_{2}$
to the injected current $\left(I_{0}=\frac{e^{2}}{h}V\right)$. Note
that this definition is slightly different from the standard Fano
factor definition due to inclusion of $\left(1-\bar{\mathcal{T}}\left(V\right)\right)$
in the denominator. This modification allows one to treat $\bar{\mathcal{T}}\approx0$
and $\bar{\mathcal{T}}\approx1$ on an equal footing and is often
employed in experimental papers.

\begin{figure*}[p]
\begin{centering}
\includegraphics[width=7cm,totalheight=5cm]{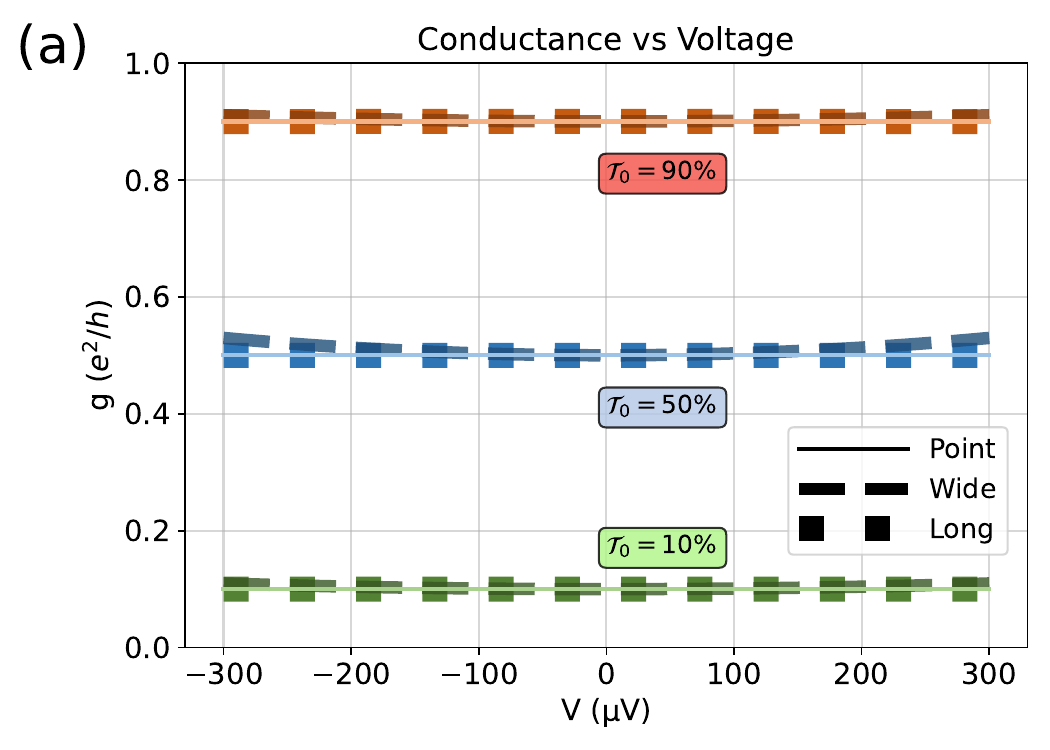}\includegraphics[width=7cm,totalheight=5cm]{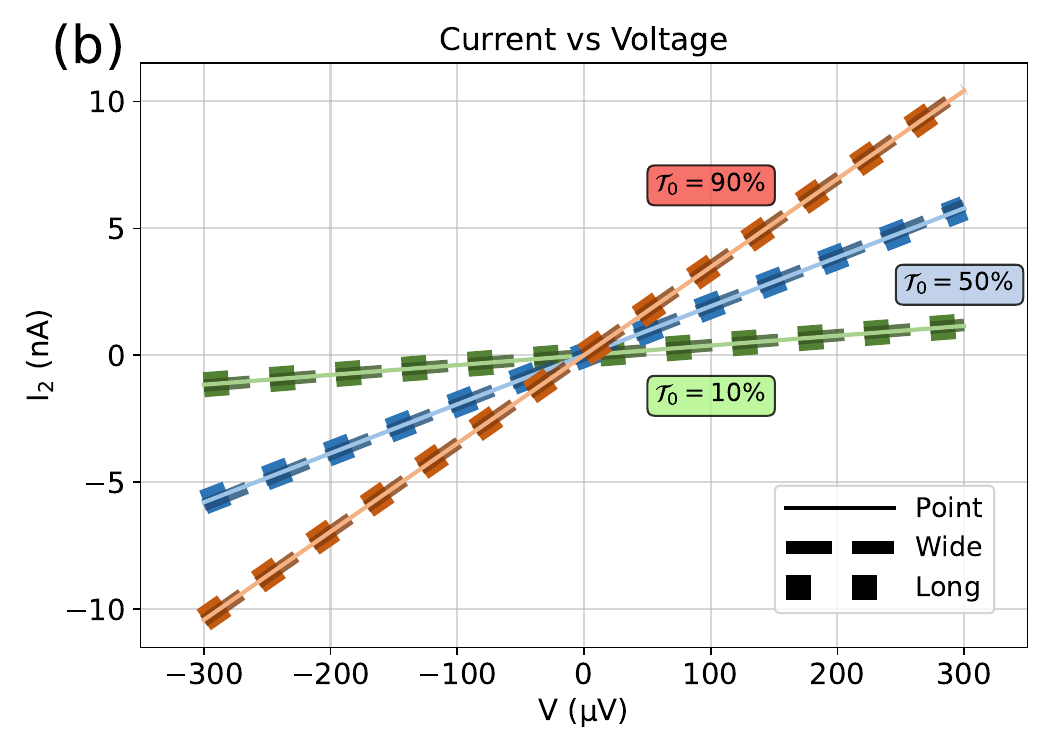}
\par\end{centering}
\begin{centering}
\includegraphics[width=14cm,totalheight=5cm]{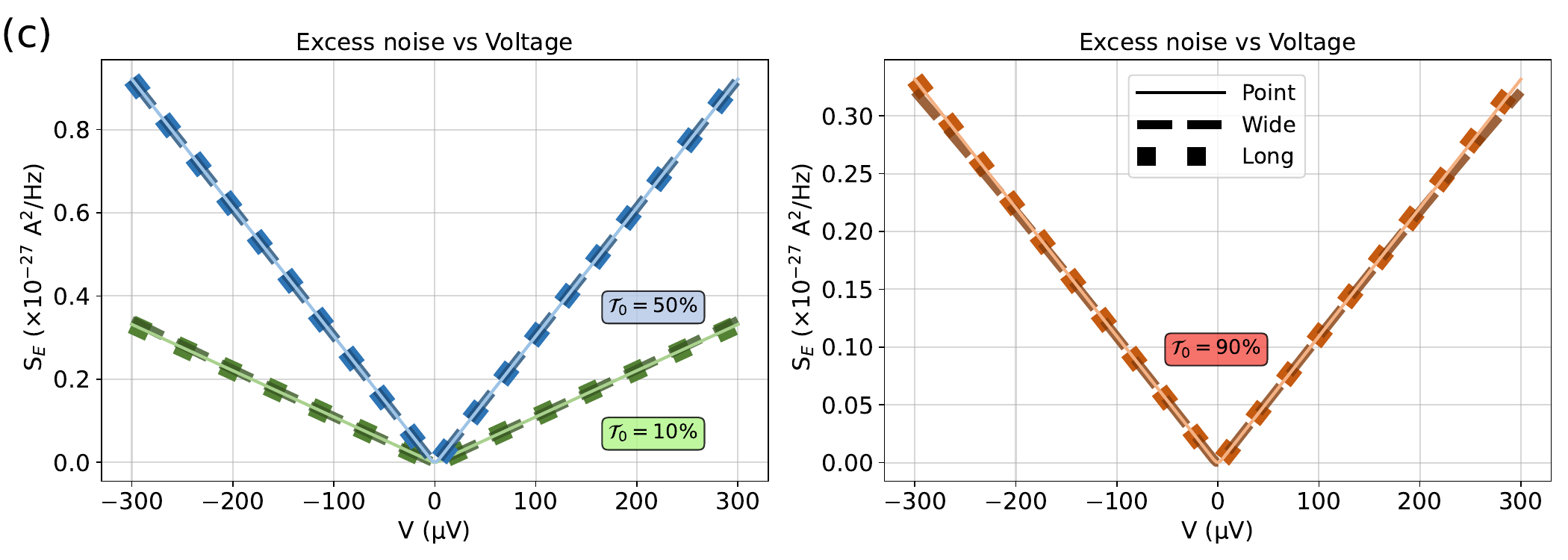}
\par\end{centering}
\begin{centering}
\includegraphics[width=18cm,totalheight=5cm]{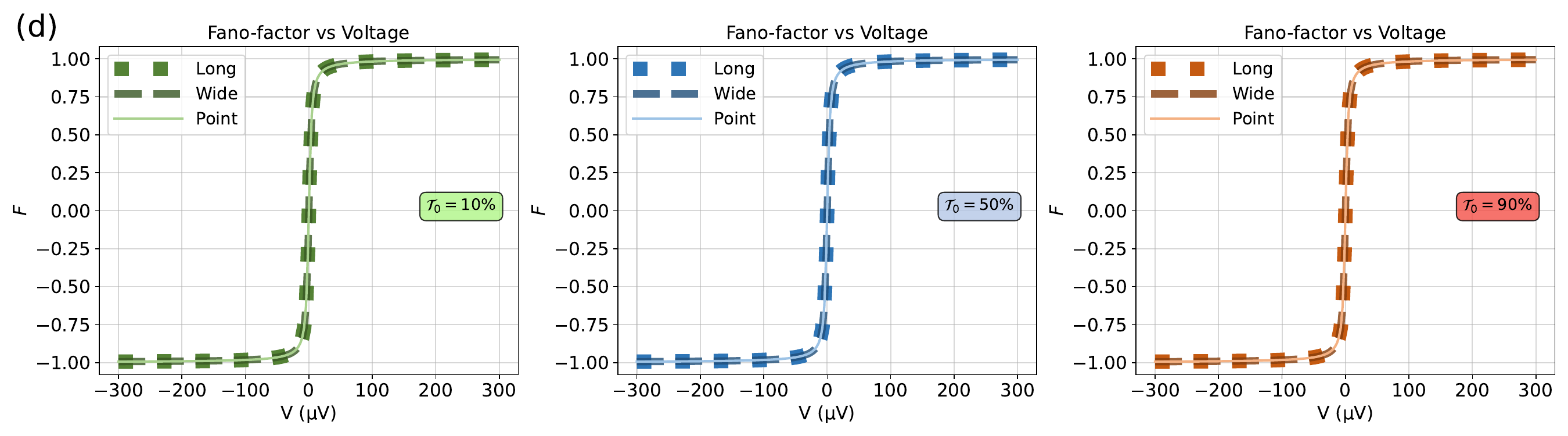}
\par\end{centering}
\caption{\label{fig:Low_voltage_transport_quantities} \textbf{Transport observables
for point-, wide-, and long-QPC models for experimentally realistic
parameters as a function of the bias voltage $V$.} (a) Differential
conductance $g$. The wide-QPC model predicts a slightly higher $g$
at large $\protect\abs V$, as compared to the point-QPC model. (b)
Current $I_{2}$. No deviation between the models is noticeable. (c)
Excess noise $S_{E}$. At low and high transmissions $\mathcal{T}_{0}$,
the noise is slightly increased in the wide-QPC model at large $\protect\abs V$.
(d) Fano-factor $F$. No deviation between the models is noticeable.}
\end{figure*}

The results for $\abs V\leq300$~$\mu$V (which is the typical voltage
range in modern experiments) are presented in Fig.~\ref{fig:Low_voltage_transport_quantities}.
Similarly to the behavior of transmission $\mathcal{T}(E)$ in Fig.~\ref{fig:T(E)_realistic_parameters}(a),
the long-QPC model is indistinguishable from the point-QPC model in
this volatge range. The wide-QPC model does exhibit noticeable yet
small deviations from the point-QPC predictions. These deviations
are can be seen in the behavior of the differential conductance $g$
and the excess noise $S_{E}$, yet are completely unnoticeable in
the behaviors of the total current $I_{2}$ and the Fano factor $F$.

\begin{figure*}[p]
\begin{centering}
\includegraphics[width=7cm,totalheight=5cm]{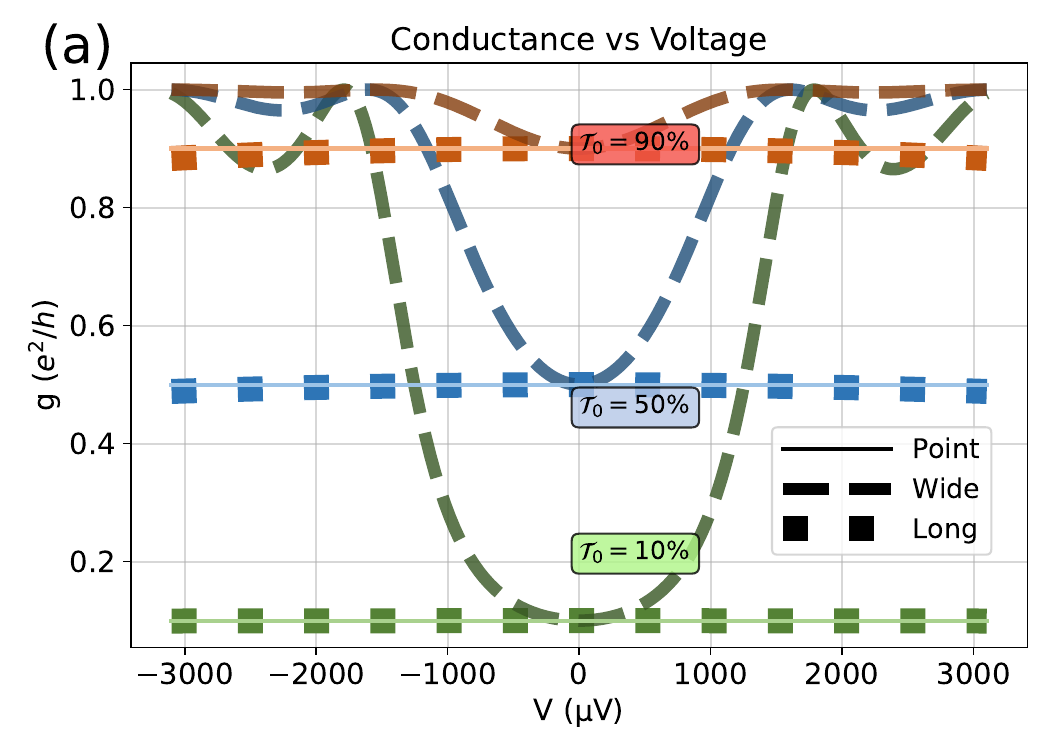}\includegraphics[width=7cm,totalheight=5cm]{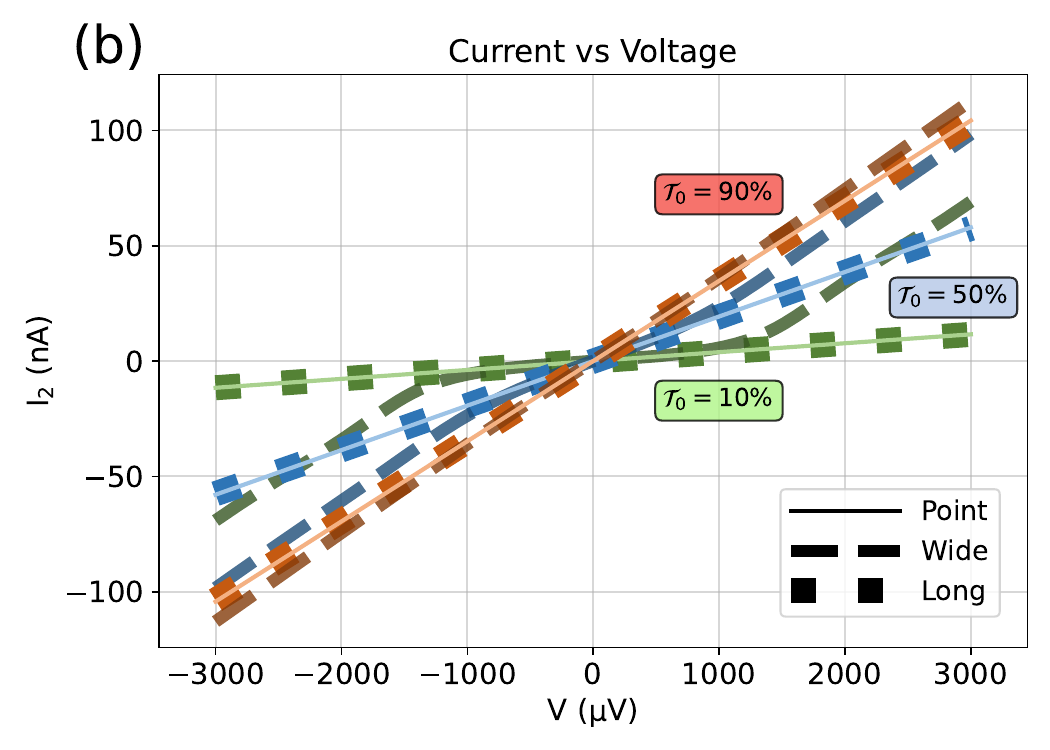}
\par\end{centering}
\begin{centering}
\includegraphics[width=14cm,totalheight=5cm]{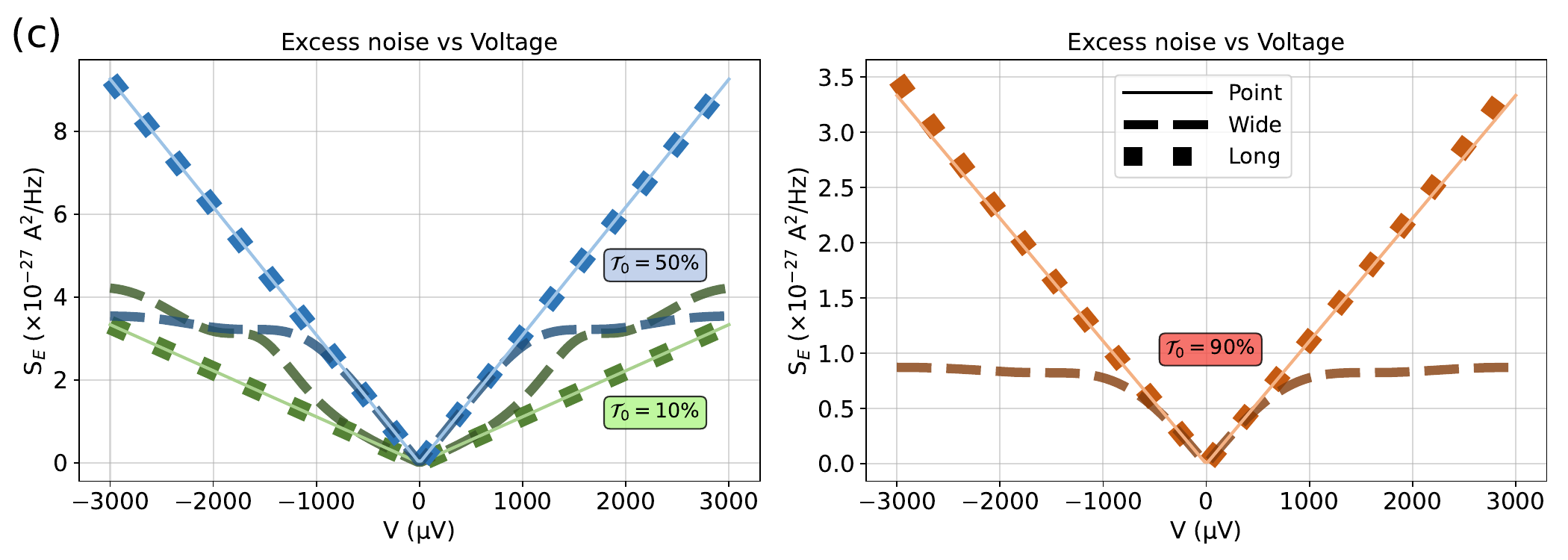}
\par\end{centering}
\begin{centering}
\includegraphics[width=18cm,totalheight=5cm]{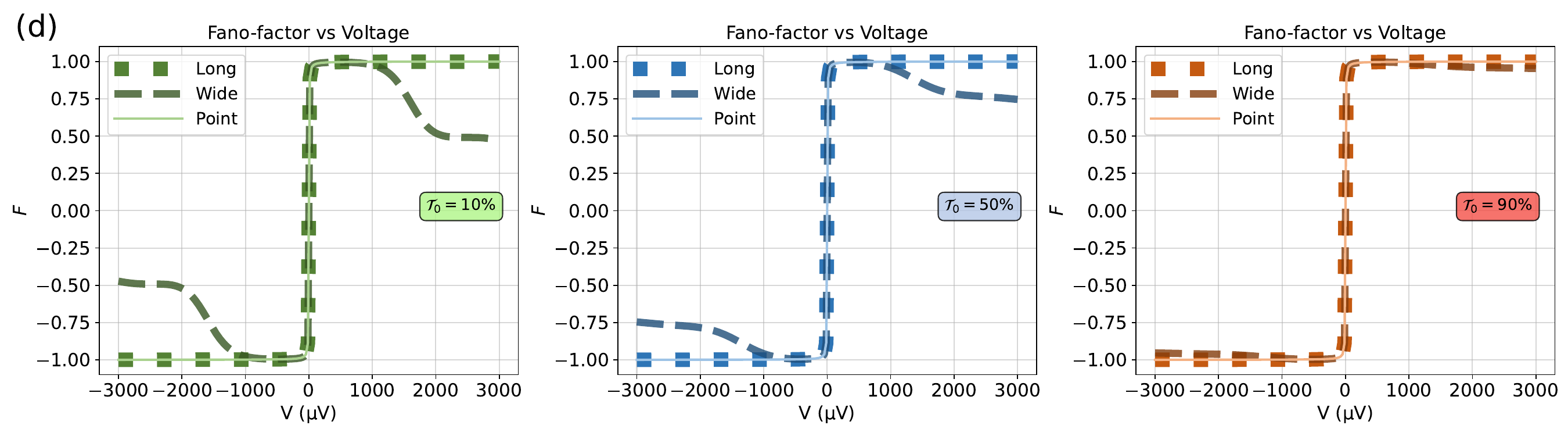}
\par\end{centering}
\caption{\label{fig: Higher_voltages_transport_quantities}\textbf{ Transport
observables for different QPC models for a voltage range larger than
typically investigated, yet experimentally reasonable.} (a) Differential
conductance $g$. (b) Current $I_{2}$. (c) Excess noise $S_{E}$.
(d) Fano-factor $F$. The long-QPC model exhibits small deviations
from the point-QPC for some observables. The deviations predicted
by the wide-QPC model are significant for all observables.}
\end{figure*}

Extending the energy range in the spirit of Fig.~\ref{fig:T(E)_realistic_parameters}(b),
we consider also the voltages up to $3000$~$\mu$V. The transport
observables in this voltage range are presented in Fig.\ref{fig: Higher_voltages_transport_quantities}.
Here the long-QPC model exhibits small deviations from the point-QPC
model for large $V$ (noticeable in $g_{2}$ and, for large $\mathcal{T}_{0}$,
in $S_{E}$, but not in $I_{2}$ or $F$). The deviations predicted
by the wide-QPC model become significant for all observables in this
voltage range (for which $e\abs V$ is still significantly smaller
than the bulk gap of $E_{c}=7.17$~meV).

\section{Conclusion}

\label{sec:conclusion}

In this work, we have considered extension of the standard point-QPC
model to account for realistic QPC geometry. Specifically, we have
introduced a model that accounts for the QPC width (wide-QPC model),
and another model that accounts for a finite distance between the
edges across the QPC (``length of the tunneling barrier'', long-QPC
model). We have shown that, in contrast to the point-QPC model, both
models predict energy dependence of the transmission function $\mathcal{T}(E)$.

We have investigated this prediction for realistic parameters in the
regime of integer ($\nu=1$) quantum Hall effect. We have shown that
the effect of the QPC width, $W$, on the behavior of the transmission
function is much more significant than that of the ``QPC length'',
$L$. For standard experimental parameters and typical ranges investigated,
the wide-QPC model does predict a noticeable energy dependence of
$\mathcal{T}(E)$ (cf.~Fig.~\ref{fig:T(E)_realistic_parameters}),
which, however, hardly translates into noticeable effects in transport
experiments (cf.~Fig.~\ref{fig:Low_voltage_transport_quantities}).
This shows that the energy dependences of $\mathcal{T}(E)$ stemming
from QPC geometries are unlikely to explain any deviations from point-QPC
predictions, at least in the integer QH regime.

Extending the range of bias voltages beyond the ones used in modern
experiments, one can see clear effects of the QPC width in transport
observables (cf.~Fig.~\ref{fig: Higher_voltages_transport_quantities}),
even if the width is of the order of the magnetic length $l_{B}$.
This extended bias voltage range still corresponds to energies significantly
below the bulk gap. Therefore, performing experiments within the extended
range and comparing their results to predictions of edge models would
be meaningful.

Finally, we note that our models do not account for the energy dependence
of the edge velocity. Introducing this aspect to the models in the
future has twofold importance. It may be necessary for modeling realistic
experiments. And from a purely theoretical point of view, it could
lift the singular behavior of the point-QPC model (see the discussion
in Sec.~\ref{sec:point-QPC-limit}).

\section*{Acknowledgments}

We acknowledge funding from the European Union\textquoteright s Horizon
2020 research and innovation program under Grant agreement No. 862683
(UltraFast Nano), from the French ANR DADDI and T-KONDO and from State
aid managed by the Agence Nationale de la Recherche under the France
2030 program, reference ANR-22-PETQ-0012 (EQUBITFLY).

This paper was prepared with the help of \href{https://lyx.org/}{LyX}
editor.

\appendix
\clearpage

\section{Derivation of the transmission function for point QPC}

\label{sec:transmission_pqpc_derivation}

In this Appendix, we provide the derivation of the point-QPC transmission
function (\ref{eq:transmission_point_qpc}). For brevity of notation,
we put $\hbar=1$ in this section.

We start with the Hamiltonian (\ref{eq:hamiltonian_point_qpc}--\ref{eq:H_tun-point}).
This yields the following equations of motion for the fields:
\begin{multline}
\partial_{t}\psi_{1}\left(x\right)=i\left[H,\psi_{1}\left(x\right)\right]\\
=-v\partial_{x}\psi_{1}\left(x\right)-i\eta\psi_{2}\left(x\right)\delta\left(x\right),\label{eq:point_qpc_left-movers}
\end{multline}
\begin{multline}
\partial_{t}\psi_{2}\left(x\right)=i\left[H,\psi_{2}\left(x\right)\right]\\
=v\partial_{x}\psi_{2}\left(x\right)-i\eta\psi_{1}\left(x\right)\delta\left(x\right).\label{eq:point_qpc_right-movers}
\end{multline}

Looking for solutions in the form
\begin{eqnarray}
\psi_{1}\left(x\right) & = & \sum_{n}\hat{a}_{n}\phi_{n,1}\left(x\right)e^{-iE_{n}t},\label{eq:solution_form_1}\\
\psi_{2}\left(x\right) & = & \sum_{n}\hat{a}_{n}\phi_{n,2}\left(x\right)e^{-iE_{n}t},\label{eq:solution_form_2}
\end{eqnarray}
where $\hat{a}_{n}$ are the Fermionic annihilation operators, we
arrive at equations for the wave functions of the system's eigenmodes:
\begin{multline}
E_{n}\phi_{n,1}\left(x\right)\\
=-iv\partial_{x}\phi_{n,1}\left(x\right)+\eta\phi_{n,2}\left(x=0\right)\delta\left(x\right),\label{eq:point_eq_of_motion_left-movers}
\end{multline}
\begin{multline}
E_{n}\phi_{n,2}\left(x\right)\\
=iv\partial_{x}\phi_{n,2}\left(x\right)+\eta\phi_{n,1}\left(x=0\right)\delta\left(x\right).\label{eq:point_eq_of_motion_right-movers}
\end{multline}
Integrating these equations around $x=0$, one finds the boundary
condition enforced by the delta function:
\begin{eqnarray}
\phi_{1}\left(0^{+}\right)-\phi_{1}\left(0^{-}\right) & = & -\frac{i\eta}{v}\phi_{2}\left(x=0\right),\label{eq:point_bd_condition_left-movers}\\
\phi_{2}\left(0^{+}\right)-\phi_{2}\left(0^{-}\right) & = & \frac{i\eta}{v}\phi_{1}\left(x=0\right).\label{eq:point_bd_condition_right-movers}
\end{eqnarray}

Equations (\ref{eq:point_eq_of_motion_left-movers}--\ref{eq:point_eq_of_motion_right-movers})
away from $x=0$ are solved by
\begin{gather}
\phi_{1}\left(x\right)=\begin{cases}
He^{\left(i\frac{Ex}{v}\right)} & \text{for }x>0\\
Ae^{\left(i\frac{Ex}{v}\right)} & \text{for }x<0
\end{cases},\label{eq:point_wf_phi_L}\\
\phi_{2}\left(x\right)=\begin{cases}
Ge^{\left(-i\frac{Ex}{v}\right)} & \text{for }x>0\\
Be^{\left(-i\frac{Ex}{v}\right)} & \text{for }x<0
\end{cases}.\label{eq:point_wf_phi_R}
\end{gather}
The boundary conditions (\ref{eq:point_bd_condition_left-movers}--\ref{eq:point_bd_condition_right-movers})
then yield
\begin{gather}
H-A=-\frac{i\eta}{v}\left(\frac{B+G}{2}\right),\\
G-B=\frac{i\eta}{v}\left(\frac{A+H}{2}\right),
\end{gather}
leading to the solution
\begin{gather}
H=-\frac{iG}{\left(\frac{v}{\eta}+\frac{\eta}{4v}\right)}+\frac{A\left(\frac{v}{\eta}-\frac{\eta}{4v}\right)}{\left(\frac{v}{\eta}+\frac{\eta}{4v}\right)},\\
B=-\frac{iA}{\left(\frac{v}{\eta}+\frac{\eta}{4v}\right)}+\frac{G\left(\frac{v}{\eta}-\frac{\eta}{4v}\right)}{\left(\frac{v}{\eta}+\frac{\eta}{4v}\right)}.
\end{gather}

\begin{figure}[t]
\begin{centering}
\includegraphics[scale=0.38]{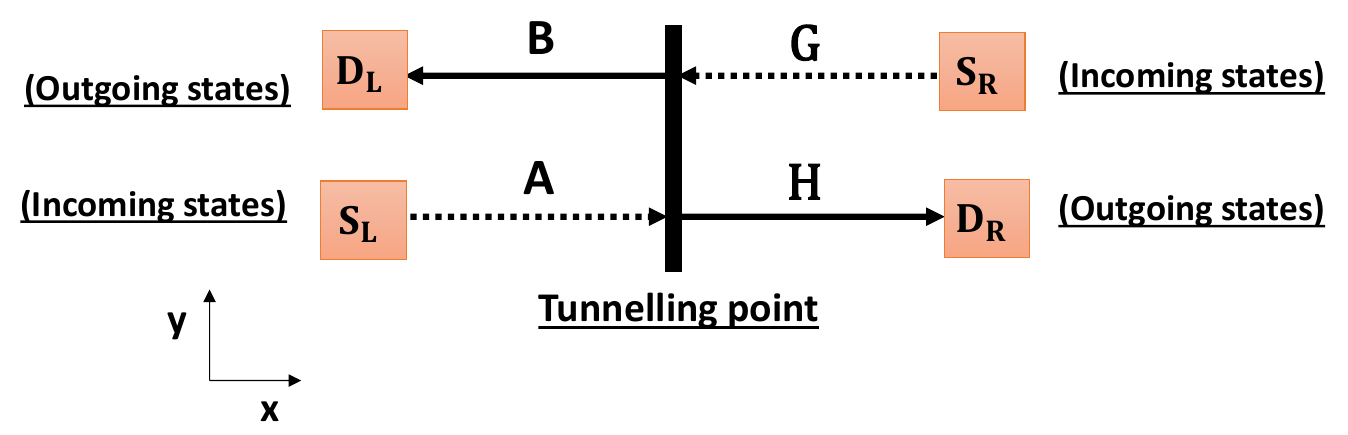}
\par\end{centering}
\caption{\textbf{\label{fig: one-point-tunnelling}Scattering point of view
on the QPC physics.}The sketch represents the QPC setup of Fig.~\ref{fig:1_QPC_experiment_scheme}
with explicit connection to the wave function in Eqs.~(\ref{eq:point_wf_phi_L}--\ref{eq:point_wf_phi_R}).
$S_{L}$ and $S_{R}$ are the sources on the left and the right side
of the QPC. The dotted lines represent the incoming edge states while
the dark lines are the outgoing edge states.}
\end{figure}

Identifying $A$ and $G$ with the amplitudes of the incoming states,
and $H$ and $B$ with those of outgoing states (cf.~Fig.~\ref{fig: one-point-tunnelling}),
one can infer the transmission and reflection amplitudes. Setting
$G=0$ ($S_{R}$ is off), one finds 
\begin{gather}
t_{L}=\frac{H}{A}=\frac{\left(\frac{v}{\eta}-\frac{\eta}{4v}\right)}{\left(\frac{v}{\eta}+\frac{\eta}{4v}\right)}=\frac{\left(1-\frac{\eta^{2}}{4v^{2}}\right)}{\left(1+\frac{\eta^{2}}{4v^{2}}\right)},\label{eq:pnt_t_L}\\
r_{L}=\frac{B}{A}=\frac{-i}{\left(\frac{v}{\eta}+\frac{\eta}{4v}\right)}.\label{eq:pnt_r_L}
\end{gather}
Similarly, setting $A=0$ ($S_{L}$ is off) leads to
\begin{gather}
t_{R}=\frac{B}{G}=\frac{\left(\frac{v}{\eta}-\frac{\eta}{4v}\right)}{\left(\frac{v}{\eta}+\frac{\eta}{4v}\right)}=\frac{\left(1-\frac{\eta^{2}}{4v^{2}}\right)}{\left(1+\frac{\eta^{2}}{4v^{2}}\right)},\label{eq:pnt_t_R}\\
r_{R}=\frac{H}{G}=\frac{-i}{\left(\frac{v}{\eta}+\frac{\eta}{4v}\right)}.\label{eq:pnt_r_R}
\end{gather}
One readily checks the unitarity of the scattering matrix $\mathcal{S}$
(\ref{eq:S-matrix}): $\mathcal{S}\mathcal{S}^{\dagger}=\mathcal{S}^{\dagger}\mathcal{S}=I$,
where $I$ is the identity matrix. The transmission function $\text{\ensuremath{\mathcal{T}_{\text{pqpc}}\left(E\right)}}=\abs{t_{L}}^{2}$
is then given by Eq.~(\ref{eq:transmission_point_qpc}).

\section{Derivation of the transmission function for wide QPC}

\label{sec:transmission_wqpc_derivation}

Here we derive the transmission function for the wide-QPC model (Fig.~\ref{fig:wide-QPC_model}),
which is defined through Eqs.~(\ref{eq:hamiltonian_wide_qpc}--\ref{eq:H_tun_wide}).
Before providing the derivation in Sec.~\ref{sec:transmission_wqpc_proper_derivation},
we discuss the properties of the extended region that features tunneling
processes. For brevity of notation, we put $\hbar=1$ in this section.

\subsection{The properties of the region with tunneling}

\label{sec:transmission_wqpc_tunnelling_region_properties}

In order to understand the properties of the extended region that
features tunneling processes, it is instructive to consider Hamiltonian
\begin{multline}
H=\int_{-\infty}^{+\infty}dx\left(-i\psi_{1}^{\dagger}v\partial_{x}\psi_{1}+i\psi_{2}^{\dagger}v\partial_{x}\psi_{2}\right)\\
+\zeta\int_{-\infty}^{+\infty}dx\left(\psi_{1}^{\dagger}\left(x\right)\psi_{2}\left(x\right)+\psi_{2}^{\dagger}\left(x\right)\psi_{1}\left(x\right)\right).\label{eq:H_infinite_tunnelling_region}
\end{multline}
Compared to Eqs.~(\ref{eq:hamiltonian_wide_qpc}--\ref{eq:H_tun_wide}),
we have taken the tunneling region to be infinite.

Since the system is translationally invariant, one can represent the
solution to the equations of motion~
\begin{align}
\partial_{t}\psi_{1}\left(x\right) & =i\left[H,\psi_{1}\left(x\right)\right]=-v\partial_{x}\psi_{1}\left(x\right)-i\zeta\psi_{2}\left(x\right),\\
\partial_{t}\psi_{2}\left(x\right) & =i\left[H,\psi_{2}\left(x\right)\right]=v\partial_{x}\psi_{2}\left(x\right)-i\zeta\psi_{1}\left(x\right),
\end{align}
in the form
\begin{align}
\psi_{1}\left(x\right) & =\sum_{k}\hat{a}_{k}A_{1}e^{ikx-iE_{k}t},\\
\psi_{2}\left(x\right) & =\sum_{k}\hat{a}_{k}A_{2}e^{ikx-iE_{k}t},
\end{align}
where $\hat{a}_{k}$ are the fermionic annihilation operators. The
equations of motion then yield
\begin{equation}
\begin{pmatrix}vk & \zeta\\
\zeta & -vk
\end{pmatrix}\begin{pmatrix}A_{1}\\
A_{2}
\end{pmatrix}=E_{k}\begin{pmatrix}A_{1}\\
A_{2}
\end{pmatrix}.
\end{equation}

\begin{figure}[H]
\centering{}\includegraphics[scale=0.5]{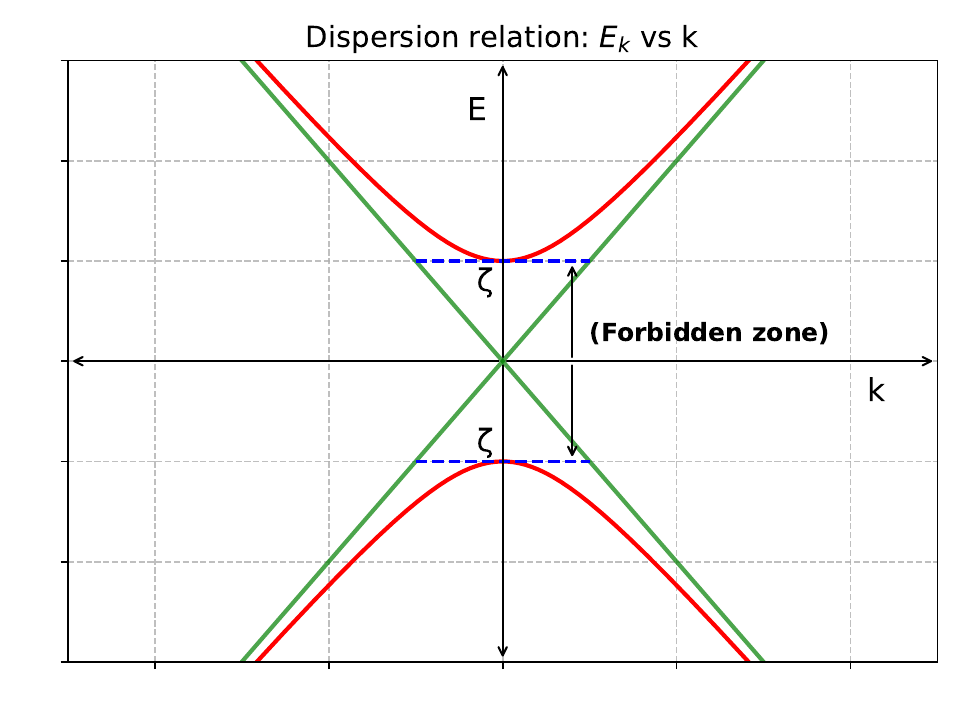}\caption{\label{fig:gapped_dispersion} Dispersion relation of the fermions
in an infinite tunneling region, cf.~Eq.~(\ref{eq:H_infinite_tunnelling_region}).
The dispersion relation of edges without tunneling $E_{k}=\pm vk$
is shown in green. In the presence of tunneling, $E_{k}=\pm\sqrt{\left(vk\right)^{2}+\zeta^{2}}$
(red), exhibiting a forbidden zone of size $2\zeta$.}
\end{figure}

The solution of this equation for eigenmodes leads to the spectrum
$E_{k}=\pm\sqrt{\left(vk\right)^{2}+\zeta^{2}}$. We thus see that
the region with tunneling features an energy gap of the size $2\zeta$
(cf.~Fig.~\ref{fig:gapped_dispersion}), which serves as an extended
tunneling barrier in the wide-QPC model.

Note that the long-QPC model, cf.~Fig.~\ref{fig:long-QPC_model}
and Eqs.~(\ref{eq:hamiltonian_long_qpc}--\ref{eq:H_tun_long}),
features the same Hamiltonian as the wide-QPC one --- up to renaming
$x\rightarrow y$, $\zeta\rightarrow\Delta$, and $W\rightarrow L$.
The interpretation of the long-QPC model is very different, though.
The tunneling region in the long-QPC model is designed to mimic the
bulk of the system. Therefore, $2\Delta$ is natural to identify with
the bulk gap. This was used to estimate the realistic model parameters
in Sec.~\ref{sec:model_comparison}.

\subsection{The transmission function}

\label{sec:transmission_wqpc_proper_derivation}

Now that we understand the effect of an extended region with tunneling,
we are in position to derive the transmission function for the wide-QPC
model. Starting from Eqs.~(\ref{eq:hamiltonian_wide_qpc}--\ref{eq:H_tun_wide}),
one obtains the equations of motion for the right-moving edge:
\begin{gather}
\partial_{t}\psi_{1}\left(x\right)=i\left[H,\psi_{1}\left(x\right)\right]\\
=\begin{cases}
-v\partial_{x}\psi_{1}\left(x\right), & x\notin\left(0,W\right),\\
-v\partial_{x}\psi_{1}\left(x\right)-i\zeta\psi_{2}\left(x\right), & x\in\left(0,W\right).
\end{cases}
\end{gather}
Similarly, for the left-moving edge:
\begin{gather}
\partial_{t}\psi_{2}\left(x\right)=i\left[H,\psi_{2}\left(x\right)\right]\\
=\begin{cases}
v\partial_{x}\psi_{2}\left(x\right), & x\notin\left(0,W\right),\\
v\partial_{x}\psi_{2}\left(x\right)-i\zeta\psi_{1}\left(x\right), & x\in\left(0,W\right).
\end{cases}
\end{gather}

Looking for the solution in the form (\ref{eq:solution_form_1}--\ref{eq:solution_form_2}),
we find 
\begin{equation}
\begin{array}{c}
\frac{d}{dx}\left(\begin{array}{c}
\phi_{1}\left(x\right)\\
\phi_{2}\left(x\right)
\end{array}\right)=\frac{i}{v}\left(\begin{array}{cc}
E & 0\\
0 & -E
\end{array}\right)\left(\begin{array}{c}
\phi_{1}\left(x\right)\\
\phi_{2}\left(x\right)
\end{array}\right)\\
\text{for }x\in\left(0,W\right),
\end{array}
\end{equation}
and
\begin{equation}
\begin{array}{c}
\frac{d}{dx}\left(\begin{array}{c}
\phi_{1}\left(x\right)\\
\phi_{2}\left(x\right)
\end{array}\right)=\frac{i}{v}\left(\begin{array}{cc}
E & -\zeta\\
\zeta & -E
\end{array}\right)\left(\begin{array}{c}
\phi_{1}\left(x\right)\\
\phi_{2}\left(x\right)
\end{array}\right)\\
\text{for }x\notin\left(0,W\right).
\end{array}
\end{equation}

\begin{widetext}

The wave function $\Phi(x)=\left(\phi_{1}(x),\phi_{2}(x)\right)^{\mathrm{T}}$
can thus be expressed in three regions:
\begin{equation}
\Phi\left(x\right)=\begin{cases}
\Phi_{\text{I}}=\left(\begin{array}{c}
Ae^{-iEx/v}\\
Be^{iEx/v}
\end{array}\right)=\left(\begin{array}{c}
Ae^{-ikx}\\
Be^{ikx}
\end{array}\right), & x<0,\\
\Phi_{\text{II}}=\frac{C}{\sqrt{2E\left(E+i\kappa v\right)}}\left(\begin{array}{c}
E+i\kappa v\\
\zeta
\end{array}\right)e^{-\kappa x}+\frac{D}{\sqrt{2E\left(E+i\kappa v\right)}}\left(\begin{array}{c}
\zeta\\
E+i\kappa v
\end{array}\right)e^{\kappa x}, & 0<x<W,\\
\Phi_{\text{III}}=\left(\begin{array}{c}
He^{-iEx/v}\\
Ge^{iEx/v}
\end{array}\right)=\left(\begin{array}{c}
He^{-ikx}\\
Ge^{ikx}
\end{array}\right), & x>W.
\end{cases}\label{eq: wide_wfs}
\end{equation}
Here $k=E/v$ and $\kappa=\sqrt{\zeta^{2}-E^{2}}/v$.

The boundary conditions at $x=0$ and $x=W$, enable expressing $A$,
$C$, $D$, and $G$ in terms of $C$ and $D$.
\begin{gather}
\Phi_{\text{I}}\left(x=0\right)=\Phi_{\text{II}}\left(x=0\right)\\
\implies\left(\begin{array}{c}
A\\
B
\end{array}\right)=\frac{C}{\sqrt{2E\left(E+i\kappa v\right)}}\left(\begin{array}{c}
E+i\kappa v\\
\zeta
\end{array}\right)+\frac{D}{\sqrt{2E\left(E+i\kappa v\right)}}\left(\begin{array}{c}
\zeta\\
E+i\kappa v
\end{array}\right),\label{eq: wide_wf_coeff_A_B}
\end{gather}
\begin{gather}
\Phi_{\text{II}}\left(x=W\right)=\Phi_{\text{III}}\left(x=W\right)\\
\implies\frac{C}{\sqrt{2E\left(E+i\kappa v\right)}}\left(\begin{array}{c}
E+i\kappa v\\
\zeta
\end{array}\right)e^{-\kappa W}+\frac{D}{\sqrt{2E\left(E+i\kappa v\right)}}\left(\begin{array}{c}
\zeta\\
E+i\kappa v
\end{array}\right)e^{\kappa W}=\left(\begin{array}{c}
He^{-ikW}\\
Ge^{ikW}
\end{array}\right).\label{eq: wide_wf_coeff_C_D}
\end{gather}
Denoting 
\begin{gather}
\alpha=E+i\kappa v,\beta=\zeta,\gamma=\sqrt{2E\left(E+i\kappa v\right)},\\
\text{and }e^{\pm\kappa W}=\cosh\left(\kappa W\right)\pm\sinh\left(\kappa W\right),\nonumber 
\end{gather}
one can write
\begin{gather}
\gamma A=\alpha C+\beta D,\,\gamma B=\alpha D+\beta C,\text{ }\label{eq: wide_wf_coeff_A_B_II}\\
\gamma H=e^{ikW}\left[\alpha Ce^{-\kappa W}+\beta De^{\kappa W}\right],\,\gamma G=e^{-ikW}\left[\beta Ce^{-\kappa W}+\alpha De^{\kappa W}\right].\label{eq: wide_wf_coeff_H_G_II}
\end{gather}
Conversely, 
\begin{equation}
C=\frac{\gamma\left(\alpha A-\beta B\right)}{\left(\alpha^{2}-\beta^{2}\right)},\quad D=\frac{\gamma\left(\alpha B-\beta A\right)}{\left(\alpha^{2}-\beta^{2}\right)},
\end{equation}
which enables one to find H and G in terms of A and B: 
\begin{gather}
H=e^{ikW}\left[\left(\cosh\left(\kappa W\right)+\left(\frac{iE}{\kappa v}\right)\sinh\left(\kappa W\right)\right)A-\left(\frac{i\zeta}{\kappa v}\right)\sinh\left(\kappa W\right)B\right],\\
G=e^{-ikW}\left[\left(\cosh\left(\kappa W\right)-\left(\frac{iE}{\kappa v}\right)\sinh\left(\kappa W\right)\right)B+\left(\frac{i\zeta}{\kappa v}\right)\sinh\left(\kappa W\right)A\right].
\end{gather}
Here we used relations
\begin{equation}
\frac{\alpha^{2}+\beta^{2}}{\alpha^{2}-\beta^{2}}=\frac{-iE}{\kappa v},\quad\frac{2\alpha\beta}{\alpha^{2}-\beta^{2}}=\frac{-i\zeta}{\kappa v}.
\end{equation}

\end{widetext}

Similarly to the consideration in Appendix~\ref{sec:transmission_pqpc_derivation},
one can find the transmission and reflection amplitudes. Setting $A=0$
(switching off $S_{L}$) leads to
\begin{gather}
t_{R}=\frac{B}{G}=\frac{\left(\kappa v\right)e^{ikW}}{\left(\kappa v\right)\cosh\left(\kappa W\right)-iE\sinh\left(\kappa W\right)},\label{eq:wide_t_R}\\
r_{R}=\frac{H}{G}=-\frac{i\zeta\sinh\left(\kappa W\right)e^{2ikW}}{\left(\kappa v\right)\cosh\left(\kappa W\right)-iE\sinh\left(\kappa W\right)}.\label{eq:wide_r_R}
\end{gather}
Setting $G=0$ (switching off $S_{R}$), one gets
\begin{gather}
t_{L}=\frac{H}{A}=\frac{\left(\kappa v\right)e^{ikW}}{\left(\kappa v\right)\cosh\left(\kappa W\right)-iE\sinh\left(\kappa W\right)},\label{eq:wide_t_L}\\
r_{L}=\frac{B}{A}=-\frac{i\zeta\sinh\left(\kappa W\right)}{\left(\kappa v\right)\cosh\left(\kappa W\right)-iE\sinh\left(\kappa W\right)}.\label{eq:wide_r_L}
\end{gather}
One checks that the unitarity of the $\mathcal{S}$-matrix is satisfied.
The transmission function 
\begin{equation}
\text{\ensuremath{\mathcal{T}_{\text{wqpc}}\left(E\right)}}=\abs{t_{L}}^{2}=\left(1+\zeta^{2}\sinh^{2}\left(\kappa W\right)/\left(\kappa v\right)^{2}\right)^{-1}
\end{equation}
can be represented as Eq.~(\ref{eq:transmission_wide_qpc}).

\section{Point limit of transmission function in conventional quantum mechanics}

\label{sec:point_limit_p^2}

\begin{figure}[H]
\begin{centering}
\includegraphics[scale=0.5]{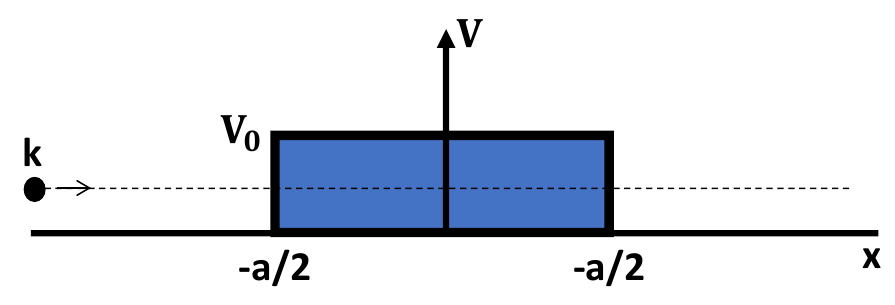}\caption{\label{fig:rectangular-potential-barrier} Rectangular potential barrier.}
\par\end{centering}
\end{figure}

Here we discuss the point limit of a rectangular barrier for a conventional
quantum particle with dispersion relation $p^{2}/(2m)$. For brevity
of notation, we put $\hbar=1$ in this section.

Consider Hamiltonian
\begin{multline}
H=\int_{-\infty}^{+\infty}dx\left(-\psi^{\dagger}\left(x\right)\frac{\partial_{x}^{2}}{2m}\psi\left(x\right)+V\left(x\right)\psi^{\dagger}\left(x\right)\psi\left(x\right)\right),\label{eq:H_conventional_dispersion_tun_barrier}
\end{multline}
where the potential 
\begin{equation}
V(x)=\begin{cases}
V_{0}, & |x|<\frac{a}{2},\\
0, & |x|>\frac{a}{2}.
\end{cases}
\end{equation}
For energies $E\leq V_{0}$ this potential represents a potential
barrier for a quantum particle, cf.~Fig.~\ref{fig:rectangular-potential-barrier}.
The probability for a particle to be transmitted through the barrier
is given by \citep[Section 1.3.11]{Band2013}
\begin{equation}
\mathcal{T}_{\mathrm{rect}}\left(E\right)=\left(1+\frac{\sinh^{2}\left(\kappa a\right)}{4}\left[\frac{\kappa^{2}+k^{2}}{k\kappa}\right]^{2}\right)^{-1},\label{eq:rect_coeff}
\end{equation}
where $k=\sqrt{2mE}$, and $\kappa=\sqrt{2m\left(V_{0}-E\right)}$.

Putting $V_{0}=\alpha/a$ and taking the limit $a\rightarrow0$ makes
the potential a delta-function barrier: $V(x)=\alpha\delta(x)$. The
transmission probability for the delta potential is given by \citep[Section 1.3.12]{Band2013}
\begin{equation}
\mathcal{T}_{\text{delta}}\left(E\right)=\left(1+\left[m\alpha/k\right]^{2}\right)^{-1}.
\end{equation}

This expression can be reproduced by taking the limit of $\mathcal{T}_{\mathrm{rect}}\left(E\right)$.
Indeed, in the limit $V_{0}=\alpha/a$, $a\rightarrow0$, 
\begin{align}
\kappa & \approx\sqrt{2mV_{0}}=\sqrt{2m\alpha/a}\rightarrow\infty,\\
\kappa a & \approx\sqrt{2m\alpha a}\rightarrow0,\\
\sinh\left(\kappa a\right) & \approx\kappa a,
\end{align}
so that
\begin{multline}
\frac{\sinh^{2}\left(\kappa a\right)}{4}\left[\frac{\kappa^{2}+k^{2}}{k\kappa}\right]^{2}\approx\frac{\left(\kappa a\right)^{2}}{4}\left[\frac{\kappa}{k}\right]^{2}\\
=\left[\frac{\kappa^{2}a}{2k}\right]^{2}=\left[\frac{m\alpha}{k}\right]^{2}.
\end{multline}
Therefore, 
\begin{equation}
\lim_{a\rightarrow0,\,V_{0}=\alpha/a}\mathcal{T}_{\mathrm{rect}}\left(E\right)=\mathcal{T}_{\text{delta}}\left(E\right).
\end{equation}

This consideration shows that in the conventional quantum-mechanical
problem with dispersion relation $p^{2}/(2m)$, the properties of
a delta-barrier can be obtained from the properties of the narrow
and high rectangular potential. Contrast this to the consideration
of Sec.~\ref{sec:point-QPC-limit}: the transmission function $\mathcal{T}_{\text{pqpc}}\left(E\right)$
obtained from the point-QPC model (which corresponds to delta-tunneling
at $x=0$) does not coincide with the point limit of the transmission
functions obtained from the wide-QPC and long-QPC models ($\mathcal{T}_{\text{wqpc}}\left(E\right)$
and $\mathcal{T}_{\text{lqpc}}\left(E\right)$). It would be interesting
to check whether taking a non-linear edge dispersion in the point-,
wide-, and long-QPC models would fix this discrepancy.

\section{The role of edge velocity}

\label{sec:role_of_edge_velocity}

In the main text (cf.~Fig.~\ref{fig:T(E)_realistic_parameters}),
we have discussed the results for the transmission function behavior
when using realistic parameters. We chose the value of velocity $v=2\times10^{4}$~m/s
for consideration there. Here we demonstrate the effect of velocity
value. In experiments, values in the range of $v=10^{4}$--$10^{6}$~m/s
have been reported \citep{McClure2009,Gurman2016,Kamata2010,Kumada2011,Kataoka2016,Nakamura2019,Nakamura2020,Weldeyesus2025}.

\begin{figure*}[p]
\begin{centering}
\includegraphics[scale=0.5]{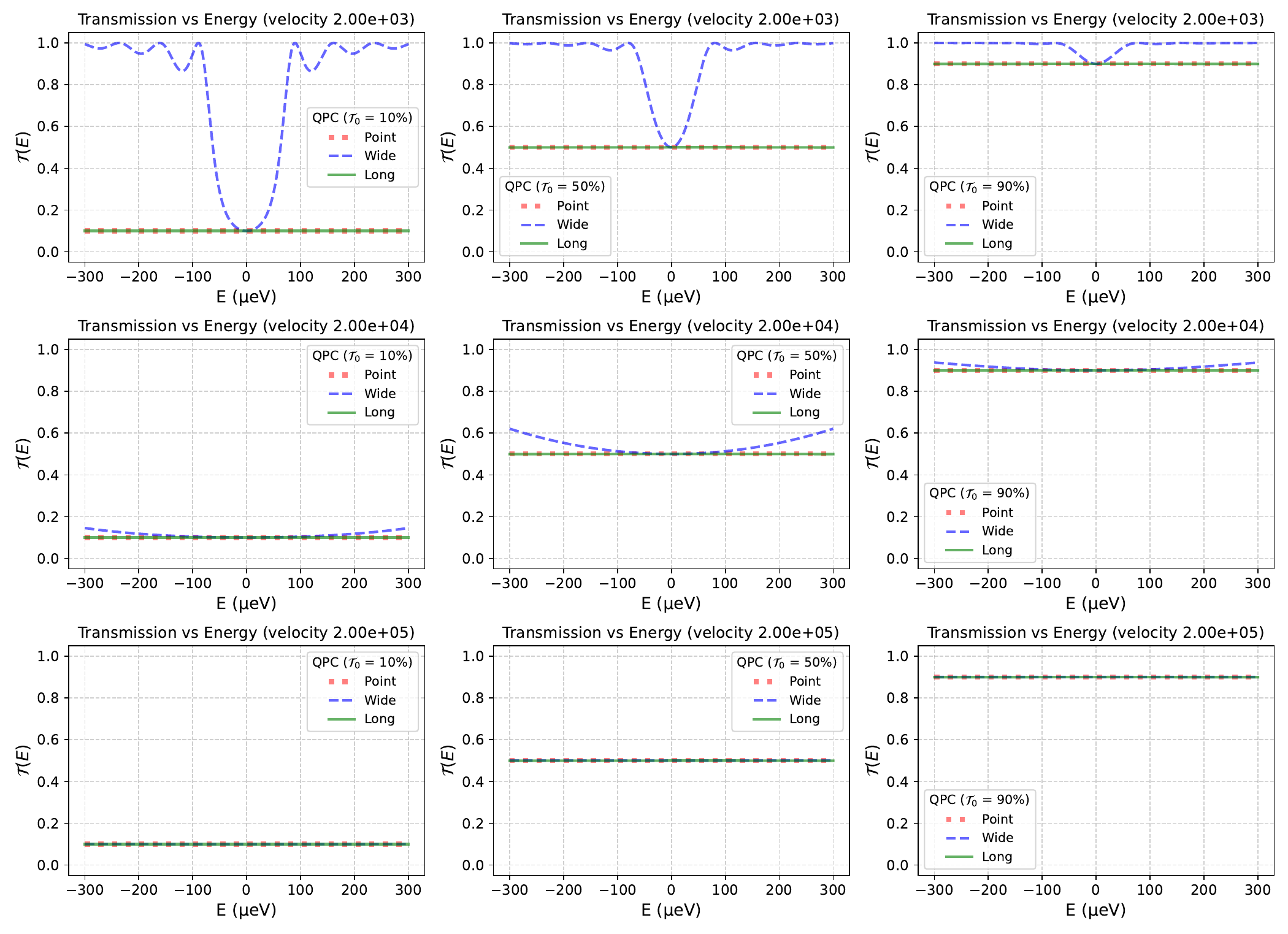}
\par\end{centering}
\caption{\label{fig: T(E)_for_diff_v} The predictions of the point-QPC, wide-QPC,
and long-QPC models for the transmission function $\mathcal{T}(E)$
several values of edge velocity $v$. The models that predict insignificant
energy dependence of $\mathcal{T}(E)$ are also not sensitive to the
velocity. By contrast, whenever the energy dependence of transmission
function is noticeable (in this energy range, wide-QPC model only),
velocity strongly affects the predictions.}
\end{figure*}

We take the same parameters as in the main text and plot the transmission
function for $v=2\times10^{3}$, $2\times10^{4}$, and $2\times10^{5}$~m/s.
The results are presented in Fig.~\ref{fig: T(E)_for_diff_v}. One
observes that velocity strongly affects the predictions of the wide-QPC
model for all values of $\mathcal{T}_{0}$. The point-QPC prediction
is unaffected as it is equal to $\mathcal{T}_{0}$ for all energies.
Similarly, the long-QPC model is also hardly affected since its transmission
function is fully determined by $\mathcal{T}_{0}$ when $E\ll\Delta$.
This shows that whenever energy dependence of the transmission function
is significant, the edge state velocity is an important factor.

\clearpage

\bibliography{bibliography}

\end{document}